\documentclass[5p,times]{elsarticle}

\usepackage{ecrc}

\volume{00}

\firstpage{1}

\journalname{XXX}

\runauth{}

\usepackage{amssymb}
\usepackage[figuresright]{rotating}
\usepackage{amsmath,amssymb,amsfonts}
\usepackage{subcaption}
\usepackage{graphicx}
\usepackage{textcomp}
\usepackage{xcolor}
\usepackage{algorithm}
\usepackage{bm,nicefrac}
\usepackage{enumitem}
\usepackage{soul, fontawesome,algpseudocode,algorithm,xspace,multirow,longtable,colortbl, lscape, pifont}
\usepackage[hyphens]{url}
\newcommand{\Eq}{Eq. } 
\newcommand{\F}{Fig. } 
\newcommand{\T}{TABLE } 

\usepackage{diagbox}
\usepackage{booktabs, caption, makecell}
\usepackage{threeparttable}
\newcommand{\jiangdraft}[1]{\textcolor{black}{#1}}

\begin{document}

\begin{frontmatter}



\dochead{}
%% e.g. \dochead{17th International Conference on Dynamical Processes in Excited States of Solids}


\title{Privacy-Utility Trades in Crowdsourced \jiangdraft{Signal Map} Obfuscation}

% \author{Jiang Zhang,
%         Lillian Clark,
%         Matthew Clark,
%         Konstantinos Psounis,
%         Peter Kairouz% <-this % stops a space
%         }
\author[label1]{Jiang Zhang}
\author[label1]{Lillian Clark}
\author[label2]{Matthew Clark}
\author[label1]{Konstantinos Psounis}
\author[label3]{Peter Kairouz}
\address[label1]{University of Southern California}
\address[label2]{The Aerospace Corporation}
\address[label3]{Google}
% \thanks{Lillian Clark, Konstantinos Psounis and Jiang Zhang are with the University of Southern California.}% 
% \thanks{Peter Kairouz is with Google.}% <-this % stops a space
% \thanks{Matthew Clark is with The Aerospace Corporation.}}

\address{}

\begin{abstract}
Cellular providers and data aggregating companies crowdsource \jiangdraft{celluar signal
strength measurements} from user devices to \jiangdraft{generate signal maps, which can be used to improve network performance}. Recognizing that this data collection may be at odds with growing awareness of privacy concerns, we consider obfuscating such data before the data leaves the mobile device. The goal is to increase privacy such that it is difficult to recover sensitive features from the obfuscated data (e.g. \jiangdraft{user ids}  and user whereabouts), while still allowing network providers to use the data for improving network services (i.e. create accurate signal maps).
To examine this privacy-utility tradeoff, we identify privacy and utility metrics and threat models suited to \jiangdraft{signal
strength measurements}. 
We then obfuscate \jiangdraft{the measurements} using several preeminent techniques, spanning differential privacy, generative adversarial privacy, and information-theoretic privacy techniques, in order to benchmark a variety of promising obfuscation approaches and provide guidance to real-world engineers who are tasked to build \jiangdraft{signal maps} that protect privacy without hurting utility.
Our evaluation results, based on multiple, diverse, real-world \jiangdraft{signal map} datasets, demonstrate the feasibility of concurrently achieving adequate privacy and utility, with obfuscation strategies which use the structure and intended use of datasets in their design, and target average-case, rather than worst-case, guarantees.
\end{abstract}


\end{frontmatter}

% \linenumbers

\section{Introduction}

%%%crowdsourcing mobile network data%%%
Network providers and data aggregating companies 
%like OpenSignal \cite{opensignal} and Tutela \cite{tutela} 
crowdsource mobile user data for a variety of reasons. This data can reveal network performance, allow for the generation of signal strength maps, inform decisions on where to deploy cell towers or sensors, and provide insight on how to improve user experience. The measurements are collected directly from user devices, via standalone mobile apps \cite{opensignal}, or measurement software development kits \cite{tutela} integrated into popular partnering apps. Providers and aggregators then sell this data to network operators, regulators, and device and equipment manufacturers. For the operators, regulators, and manufacturers, this crowdsourced data offers clear value for network planning. For the user, contributing data can in turn be useful, given that it leads to better network performance. However, participation also raises legitimate privacy concerns. 

%%%privacy threat%%%
For example, some cellular providers have allegedly been selling their users' real-time location data to credit agencies, bail bondsmen, and other third parties \cite{att_lawsuit}. Furthermore, while these measurements are assumed to be sparse in space and time and over thousands of users, previous work has shown that identities are inferable from anonymized data \cite{personal_discovery}.

%%%obfuscation and privacy-utility tradeoff%%%
In recent years, privacy issues have come to the front of news, politics, and public opinion \cite{nature_privacy, nytimes, ibm} and pioneering privacy laws have been enacted \cite{URL_GDPR,URL_CCPA}. To protect user privacy, a plethora of data masking, or obfuscating, schemes have been proposed, see, for example, \cite{privacycrowdsensing}. However, by obfuscating the original data for the sake of privacy, data can no longer provide the exact insights it once could, sacrificing data utility for privacy \cite{infotheor_util}. 

%%%introduction of application and motivation: benchamrking for real world engineers%%%
In this work we examine the privacy-utility tradeoff in the context of \jiangdraft{cellular signal strength measurements}, focusing on device-level obfuscation where \jiangdraft{the measurement} is obfuscated, or privatized, before it leaves the user's phone.
The goal is to increase privacy such that it is difficult to recover sensitive features from \jiangdraft{the obfuscated measurements}, including user ids and whereabouts, while still allowing network providers to use the measurement for improving network services, i.e. create accurate signal maps.
To examine this privacy-utility tradeoff, we identify privacy and utility metrics and threat models suited to the \jiangdraft{signal map} application at hand. 
We then obfuscate \jiangdraft{the measurements} using a number of promising approaches at the forefront of privacy research, in order to 
benchmark them and provide guidance to real-world engineers who are tasked with 
building \jiangdraft{signal maps} that provide (some) privacy while maintaining (adequate) utility.
To evaluate the different approaches, we use multiple, diverse, real-world \jiangdraft{signal map} datasets to ensure real world applicability of our findings.

%%%obfuscation techniques considered%%%
%Specifically, 
We implement four strategies for obfuscating \jiangdraft{signal strength measurements} to assess and compare their application-specific performance, selecting preeminent methods from the literature that span a range of complexities and privacy guarantees. %against different adversary threat models. 
Specifically, the first is a noise-adding privatizer, which adds independent, identically distributed Gaussian noise across the features of the data. Albeit simple, this scheme provides intuition into the privacy-utility tradeoff via the choice of how much noise to add. The second is based on differential privacy (DP) \cite{dp}, a leading approach to data obfuscation which provides probabilistic worst-case guarantees against any arbitrary adversary, including one with unlimited resources and access to side-information. %through a differentially private mechanism. 
In this work we apply the popular local Gaussian mechanism \cite{dp}, as well as the recent Truncated Laplacian Mechanism \cite{geng2018privacy}.
The third leverages the idea of generative adversarial networks to allow a data-driven method of learning an obfuscation scheme. This method, which is referred to as generative adversarial privacy (GAP) \cite{gap}, positions a privatizer and an adversary, both modeled as neural networks, against each other. The privatizer learns to obfuscate the data such that the adversary cannot infer sensitive features, and the adversary simultaneously learns to infer sensitive features. 
While this method cannot offer the formal worst-case guarantees of the differentially private methods, the learning approach offers the potential to leverage structure in the data set and take advantage of the specific utility objectives in the network.
The fourth strategy is motivated by an information-theoretic treatment of the problem. Considering mutual information as a convex metric for privacy performance, we frame a formal optimization problem as finding the obfuscation strategy which maximizes privacy subject to a constraint on utility. This approach maximizes user privacy in an average sense, but sacrifices the worst-case guarantees offered by the deferentially private methods. Section \ref{sec:privatizer} discusses these privatizers in more detail. 

%%%datasets used%%%
We analyze the performance of each of these privatizers using 
three, diverse, real-world \jiangdraft{signal map} datasets. The first one is collected from cellular users over a seven month period in the city of Chania, Greece \cite{manos}. The second one is collected over a period of four months by Android smartphones in the University of California Irvine campus \cite{emmanouil2017using}. The last one is sampled from the Radiocell dataset \cite{URL_RADIOCELL}, one of the largest publicly available datasets with millions of measurements from nearly one million macrocells around the world. The sample we work with contains \jiangdraft{signal strength} measurements from hundreds of users over a one year period in UK's countryside. Section \ref{subsec:userdata} discusses these datasets in detail.

%%%time to talk about metrics and threat models%%%
An important aspect of our study is to identify privacy and utility metrics (Section \ref{sec:metric}) as well as threat models (Section \ref{subsec:threatmodel}) suited to \jiangdraft{signal map application}.
We assess our obfuscation schemes against specific adversaries, modeled as neural networks (Section \ref{subsec:adversary} discusses adversary models in detail), which estimate %the user ID and unobfuscated user location. 
private user information from observing obfuscated data, and we take the adversary's estimation performance as a practical, application-specific privacy metric. We also consider more robust privacy guarantees, such as DP, which is not dependent on any specific adversary implementation.
With respect to utility, we consider two metrics. 
First, we consider a received signal strength (RSS) model which accurately predicts signal 
maps when trained with unobfuscated data.  We train this model with the obfuscated data.
Then, we use as an application-specific utility metric the $L_1$ distance between the parameters of the RSS model trained with obfuscated versus unbofuscated data.
%application-specific metrics, namely the generation of signal maps and user location maps, which are analogous to heat maps of supply and demand respectively. 
%If maps generated by the obfuscated data closely resemble those generated by the input data, utility is high.
As a general utility metric, we use the overall assessment of data distortion. 
This serves as a proxy for utility under a wide variety of other potential mobile data applications.

The main contributions of this work are: (1) a framework for formalizing privacy and utility for crowdsourced \jiangdraft{signal map application}, (2) a systematic exploration of the parameterized privacy-utility tradeoff \jiangdraft{when obfuscating signal strength measurements}, 
(3) an analysis and comparison of four obfuscation schemes, 
and (4) an evaluation of the feasibility of achieving different notions of privacy in the \jiangdraft{signal map application}. Our findings show that while local DP provides privacy guarantees under worst-case conditions, it comes with a substantial cost in utility. GAP and IT privacy can offer significantly improved privacy-utility tradeoff by sacrificing worst-case guarantees and incorporating application-specific context, such as structure in the datasets and network objectives.

In the next section, we briefly discuss relevant work in privacy, especially as it relates to mobile network data. Section \ref{sec:model} describes our system model, including
the three real-world datasets that we use in our evaluation, the threat models we consider, the privatizer and adversary model we implement, and the service provider model we consider. Section \ref{sec:metric} rigorously defines our privacy and utility metrics. Section \ref{sec:privatizer} presents each of the four obfuscation schemes. In Section \ref{sec:result}, we evaluate and compare these schemes, and analyze our results.  We discuss the limitations and future works in Section \ref{sec:limitation}, and present our conclusions in Section \ref{sec:conclusion}.
\section{Related Work}
%We first discuss privacy mechanisms and theoretical studies on privacy-utility trades. We then discuss prior work specific to our application.
% \subsection{Signal Maps} \jiangdraft{Prior works have studied the generation of signal maps based on crowdsourced measurements \cite{fida2017zipweave,chakraborty2017specsense}. However, the above works have not investigated user privacy during signal map generation.}
\subsection{Privacy mechanisms}  

Differential privacy (DP) \cite{dworkdp, dp, dpresults} is a mathematically rigorous definition of privacy which is useful for quantifying privacy loss and designing randomized algorithms with privacy guarantees. Motivated by statistical disclosure control, or providing accurate statistics while protecting individual survey respondents, DP approaches the problem of releasing coarse-grained information while keeping fine-grained details private. This popular approach to data privatization is studied under the local \cite{localdp} and global models. The global model assumes a trusted data analyst has access to the dataset and wants to release queries computed on it in a privacy-preserving fashion. The local model assumes the absence of a trusted server, thus the data is randomized prior to aggregation. This work applies the local model of differential privacy. 

Generative Adversarial Privacy (GAP) \cite{gap, cagap} offers an alternative to noise-adding mechanisms in that it is context-aware, meaning it takes the dataset statistics into account. GAP learns from the dataset without the need to explicitly model the underlying distribution. Leveraging recent advancements in generative adversarial networks (GANs) \cite{gan1, gan2, cgan}, GAP allows the privatizer to learn obfuscation schemes from the dataset itself. Like the generator and discriminator in a GAN, the privatizer and adversary optimize a minimax game. 

Information-theoretic (IT) privacy \cite{infotheor_util, info_theo, infotheoretic} provides an alternative in which privacy metrics are motivated by concepts from information theory. For example, mutual information \cite{mutualinformation} is the measure of how much one random variable tells us about another. Obfuscation schemes which minimize mutual information intuitively provide privacy. Unlike DP which provides guarantees on worst-case privacy, mutual information is an expectation, i.e. provides guarantees on average privacy.

\subsection{Theoretical studies of privacy-utility trades}

Previous work has analyzed distortion in the context of DP \cite{dp_distortion} or attempted to minimize the utility loss incurred by DP \cite{dp_lopub}. 
Previous work in GAP maximizes privacy subject to a constraint on distortion \cite{gap}. 
Additionally, previous 
IT privacy metrics have been considered in the context of theoretically motivated utility metrics \cite{ITutilitymetrics}.
In contrast, in this work we consider utility metrics beyond distortion which are specific
to our application and are both more intuitive and relevant for mobile network data. 
We also formally compare the performance of context-free (Gaussian noise-adding, local DP) and context-aware (GAP, IT) approaches in the context of our application. 
%
%Previous work has analyzed distortion in the context of DP \cite{dp_distortion} or attempted to minimize the utility loss incurred by DP \cite{dp_lopub}.
%In this work we further examine parameterizing this privacy-utility tradeoff in the context of mobile network data, and utility metrics specific to this application.
%
%Previously, information-theoretic privacy metrics have been considered in the context of theoretically motivated utility metrics \cite{ITutilitymetrics}. In this work we evaluate obfuscation schemes motivated by information theory against application-specific utility metrics which are both more intuitive and relevant for mobile network data.
%
%Previous work in generative adversarial privacy maximizes privacy subject to a constraint on distortion. In this work we consider utility loss functions beyond distortion, and maximize utility and privacy jointly. We formally compare the performance of GAP with other context-free (Gaussian noise-adding, local DP) and context-aware (information-theoretic) %approaches.

Prior theoretical studies on the privacy-utility tradeoff include \cite{infotheor_util, PU_trade, pu_ghosh}. The authors of \cite{infotheor_util} formally define an analytical model for trading equivocation (a privacy metric based on Shannon entropy) and distortion (a utility metric which could be Euclidean distance, Hamming distortion, Kullback-Leibler divergence, etc). This model is designed for ``universal" metrics, but is not generalized for non-i.i.d. datasets or datasets lacking strong structural properties. A so called geometric mechanism is presented in \cite{pu_ghosh} as a utility-maximizing alternative to the Laplace or Gaussian mechanisms typically used in differential privacy, where utility is the expected loss of any symmetric, monotonic loss function. In \cite{PU_trade}, the authors define a bound on the information-theoretic min-entropy leakage of $\epsilon$-differential privacy, and a bound on utility (where utility is roughly the number of differing dataset entries).
Our work uniquely examines this tradeoff for all of these approaches in the unifying context of a single application, allowing us to present additional insight.

%**********
\subsection{Prior work on mobile network data privacy}
\label{subsec:prior_work_net}
Previous work on privacy in mobile network data has considered strategic sampling, distribution modeling, and noise addition as obfuscation strategies. In \cite{privacycrowdsensing}, the authors exploit compressive sensing techniques to sample and compress received signal strength values in a privacy-preserving RSS map generation scheme. While privacy is gained in sampling and compression, the authors of \cite{privacycrowdsensing} do not take a formal approach to quantifying privacy. In \cite{distributednoise}, distributed algorithms for Gaussian and exponential noise addition are explored in a crowdsourced data setting. Local differential privacy is applied to the user-tracking problem for indoor positioning systems in \cite{mobile_dp}. The authors of \cite{syntheticdata} present a relaxed version of differential privacy, probabilistic DP, which accounts for certain worst-case privacy scenarios being highly unlikely. They apply this to the generation of synthetic data which maps daily commutes. In \cite{jin2016inception}, a novel privacy-preserving incentive mechanism is proposed for mobile crowd sensing, where the authors employed DP to perturb aggregated data. In each of \cite{privacycrowdsensing,distributednoise, mobile_dp, syntheticdata,jin2016inception}, utility is not rigorously considered. Our work takes a formal approach to both privacy and utility. 

In recent years, researchers have grown interested in studying the privacy-utility tradeoff in mobile network applications. Shokri et al. in \cite{shokri2012protecting} propose an optimal strategy against location attack based on Stackelberg Bayesian game theory, which provide the best location privacy while satisfying the user’s service
quality requirements. Bordenabe et al. in \cite{bordenabe2014optimal} formulate the tradeoff optimization problem between
geo-indistinguishability and quality of service, and propose a method based on linear optimization to solve this problem. 
Chen et al. in \cite{chen2021optimized} design a differentially private obfuscation scheme based on reinforcement learning to optimize privacy budget allocation for each location in vehicle trajectory obfuscation, which can balance geolocation obfuscation and semantic security and thus results in better privacy-utility tradeoff.
In \cite{boukoros2019lack}, the authors design novel privacy and utility metrics for location privacy, and perform large-scale evaluation and analysis of several existing location-privacy preserving mechanisms. 
In \cite{KIM2021102464}, the authors provide a survey of DP-based obfuscation approaches for location privacy protection and compare the privacy-utility tradeoff performance of these approaches.
However, these works only focus on location privacy without considering other privacy metrics for mobile user data. Moreover, the proposed approaches in \cite{shokri2012protecting} and  \cite{bordenabe2014optimal} are based on linear programming and discrete locations which cannot be easily applied under our threat models (continue locations and non-linear adversary).
The mechanism proposed by \cite{chen2021optimized} does not formally optimize the privacy-utility tradeoff during trajectory obfuscation.
\cite{boukoros2019lack} is an empirical study with no formal analysis or obfuscation schemes that formally consider both privacy and utility in their design, and \cite{KIM2021102464} only surveys DP-based obfuscation approaches without considering other obfuscation schemes.
%*****They only consider noise adding DP which does not account for utility******

In \cite{gursoy2018differentially}, the authors propose a novel framework, DP-star, for publishing trajectory data with differential privacy guarantees, while preserving high utility.
In \cite{adatrace} the authors present AdaTrace, a utility-aware location trace synthesizer which provides a differential privacy guarantee and inference attack resilience. This work is closely related to ours in that the authors employ both learning and noise-adding to generate datasets which they evaluate for statistical utility, and analyze how the choice of privacy parameter effects utility.
In \cite{messaoud2018privacy}, the authors proposed a privacy-preserving and utility-aware mechanism based on mutual information optimization, with application to the data uploading phase in participatory sensing. 
Zhang et al. in \cite{zhang2020aggregation} also propose an information theoretic approach based on mutual information optimization, which protects the user's location privacy while satisfying the user's utility constraints when releasing location aggregates.
However, these works only consider the database-level threat model during dataset publishing, which requires a trust-worthy third party to distort data before release, and they cannot be directly applied to the device-level obfuscation in our application.

In \cite{liu2019privacy} and \cite{Raval2019OlympusSP}, GANs are leveraged to achieve utility-aware obfuscation of mobile sensor data. However, these works focus on obfuscating image sensor data to reduce sensitive information leakage in mobile apps, where both the dataset structure and threat model are different from ours. Moreover, \cite{Raval2019OlympusSP} does not compare GANs with other formal obfuscation schemes, and \cite{liu2019privacy} does not compare against obfuscation schemes that formally consider both privacy and utility in their design.
%Note that their methodologies are the same as our GAP privatizer.

While the advantages and disadvantages of a range of obfuscation methods are to some extent known in principle from prior work, how they perform and compare in \jiangdraft{signal map} application is unclear. In this work, we implement representative obfuscation schemes based on preeminent approaches, apply them to the important real-world application of generating signal maps via crowd-sourcing, and compare their performance. Our performance results can serve as benchmarks, offering insights about how to design real-world systems to generate accurate signal maps while protecting user privacy.

%In this work, we further examine the privacy-utility tradeoff. However, our work defines both general and application-specific metrics and considers both theoretical formulations and empirical measures of privacy and utility.
\section{System Model}
\label{sec:model}
%We focus on mobile networks involving users carrying radio frequency devices, where users intermittently report data to a service provider. Each data report is composed of a set of features, such as location and received signal strength (RSS). The service provider aggregates the reports to inform network planning decisions. For example, reports on received signal strength (RSS) can be applied to identify coverage gaps, while information on typical user density can be used to inform required capacity, e.g., of new base stations. The reported data contains information that the users would deem private. For example, location and user ID can be combined and pose a significant threat to any user, particularly if multiple reports are collected over time and an adversary can recreate a trace of the user's behavior. 

\begin{figure}
    \centering
    \includegraphics[width=\linewidth]{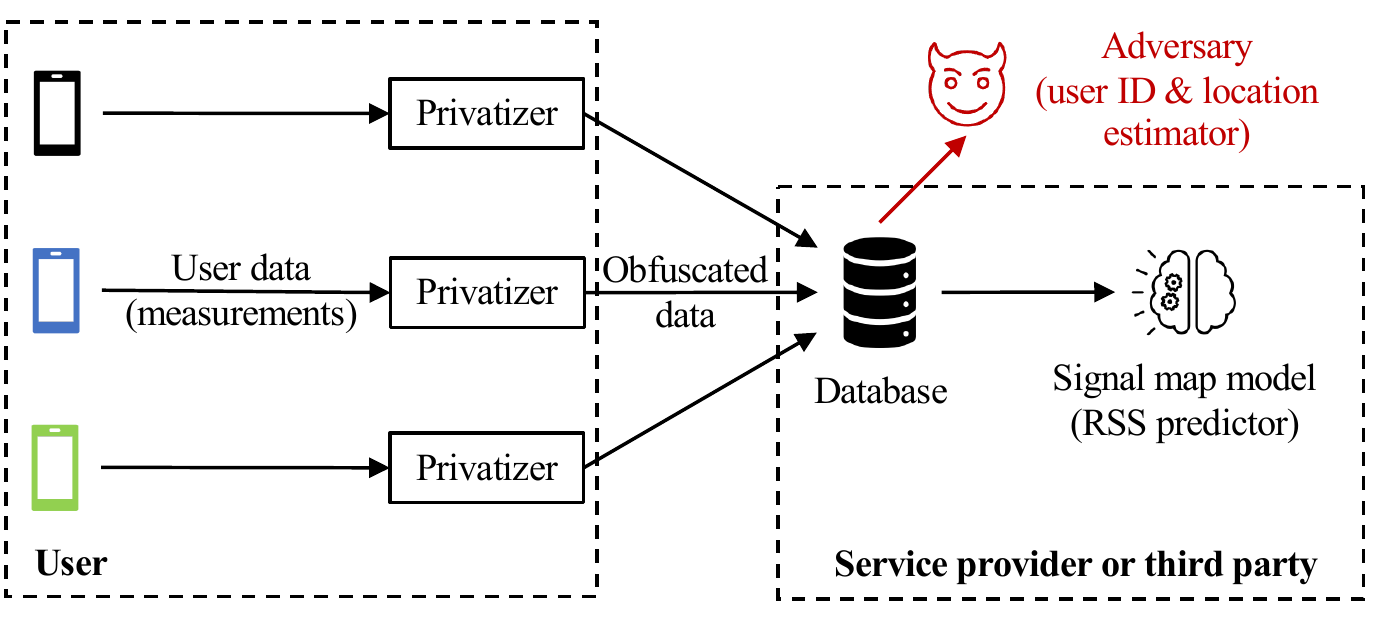}
    \caption{Overview of system model. 1) Users collect measurement data and obfuscate it before uploading it to the service provider; 2) The service provider or third party aggregates obfuscated user data to train a signal map model, i.e. RSS prediction model; 3) The adversary has access to the obfuscated user data and uses it to estimate the user ID and locations.}
    \label{fig:sys_model}
\end{figure}

\F\ref{fig:sys_model} illustrates the system model we consider, which involves mobile users, a service provider or a third party, and an adversary. User devices record network measurement data and transmit it to a service-provider or third-party server. Since the reported data contains information that the users may deem private (e.g. user location, see Section \ref{subsec:userdata}), users apply device-level privatizers to obfuscate their data locally before uploading them to the server (see Section \ref{subsec:pri}).
The goal of the service provider is to train a RSS model based on the aggregated obfuscated user measurement data, which can be used to generate signal maps and thus guide network planning and operation \cite{taufique2017planning}  (see Section \ref{subsec:service}). Finally, an adversary with access to the obfuscated data estimates the whereabouts of users, by estimating the user ID and location corresponding to the incoming measurements (see Section \ref{subsec:adversary}). Note 
that we assume the adversary has access to the obfuscated data as it arrives at the server, but no side information that directly reveals 
the identity of users  (see Section \ref{subsec:threatmodel} for a detailed description of the threat model).

% Note: adversary has access to database of service provider, observe the updates to database, but do not have access to connection information that would reveal the identity of users 

\subsection{User Data}
\label{subsec:userdata}
%\textcolor{blue}{The specific features we implement in our model are intended to be representative and specific enough for quantitative analysis, recognizing the intractability of enumerating all possible features that could be relevant across mobile data applications. Our representative feature set is driven by the available datasets we will use later to produce experimental results.} 
We use three real-world datasets collected from different countries and over different time periods to evaluate the performance of our privatization schemes under different environments and user behaviors, and thus make our findings more conclusive.

The first dataset is taken from users in Chania, Greece, and will be referred to as the Chania dataset, which contains measurements from nine users over seven months in 2014. The nine users are mobile device owners who carry their devices with them throughout the day collecting measurements. Each measurement contains 24 features: device address, timestamp (to the second), received signal strength (RSS) in dBm, latitude, longitude, cellID identifying the base station, downlink carrier frequency, uplink carrier frequency, mobile network code, etc. 
%For completeness, no features were removed from the dataset. For convenience, we convert the device addresses into user IDs 0-8.

The second dataset contains measurements from seven users over four months in the University of California Irvine (UCI) campus in 2017, and will be referred to as the UCI dataset. Each measurement consists of 15 features including latitude, longitude, reference signal received power (RSRP) in dBm, reference signal reference quality (RSRQ) in dBm, timestamp. deviceID, cellID, etc.

The third dataset is collected by Radiocell.org \cite{URL_RADIOCELL}, which has been crowdsourcing wireless network measurements from world-wide mobile users since 2009. It is the largest open-source mobile network dataset we can have access to. We sample about 0.5 million measurements from 219 mobile users in UK, 2019\footnote{We choose UK since most of the collected measurements in 2019 come from mobile users in UK (10 million measurements in total). To %further reduce the dataset size, 
limit computational complexity, we select three cells containing the largest amount of data.}, and refer to it as Radiocell dataset. Each measurement has 23 features including latitude, longitude, altitude, speed, signal strength (SS) in dBm, country code, mobile network code, etc.

The most relevant features to this paper are tabulated in Table \ref{table:features} along with an indicator of their sensitivity. User ID and location are assumed sensitive features (private), whereas RSS/RSRP and others are not sensitive (public). 
% \renewcommand{\arraystretch}{1.3}
% \begin{table}[h]
% \caption{Dataset Features}
% \centering
% \begin{tabular}{c c c}
% \hline\hline
% Feature & Sensitivity & Variable \\ %[0.5ex]
% \hline
% user ID & private & $u$ \\
% latitude & private & $x_1$ \\
% longitude & private & $x_2$ \\
% RSS & public & $x_3$ \\
% others & public & $x_{i,i>3}$\\
% \hline
% \end{tabular}
% \label{table:features}
% \end{table}
% \renewcommand{\arraystretch}{1}
\renewcommand{\arraystretch}{1.3}
\begin{table}[h]
\caption{Dataset Features}
\centering
\footnotesize
\begin{tabular}{c|ccccc}
\hline\hline
Features & User ID & Latitude & Longitude & RSS & Others \\\hline
Sensitivity & Private & Private & Private & Public & Public \\\hline
Variable & $u$ & $x_1$ & $x_2$ & $x_3$ & $x_{j,\mbox{ }j>3}$\\
\hline
\end{tabular}
\label{table:features}
\end{table}
\renewcommand{\arraystretch}{1}
\begin{figure}{
    \centering{
    \includegraphics[width=\columnwidth]{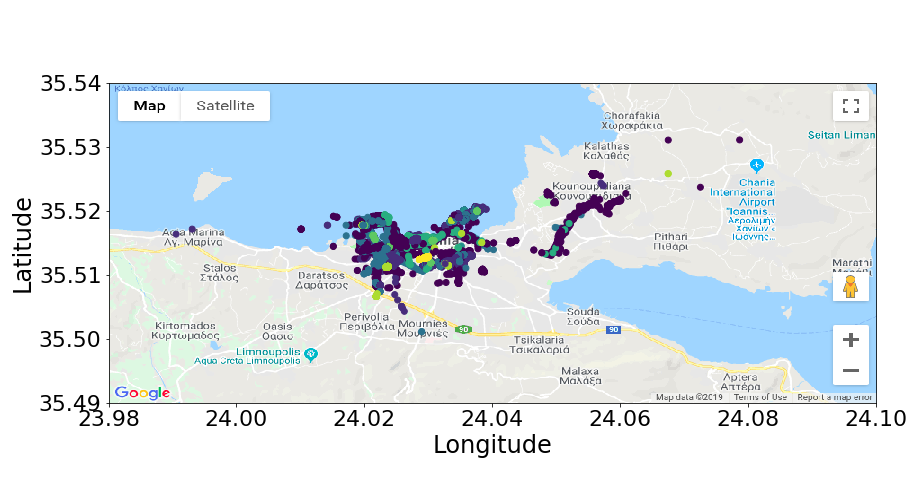}}
    \caption{Chania Dataset (colored by user)}
    \label{fig:chania_overlay}}
\end{figure}

For visualization purposes, we have plotted the data of the first dataset over the geographic region in \F\ref{fig:chania_overlay}. The colors indicate user ID, and it is apparent that one cannot easily assume user ID based on location alone.

\subsection{Threat Model}
\label{subsec:threatmodel}
% \begin{table}[h]
%     \centering
%     {\color{red}\begin{tabular}{p{4.6cm}|c|c}
%     \hline\hline
%          \backslashbox{Properties}{Threat Models} & Typical & Worst-case \\ \hline
%          Adversary computational resources & limited & unlimited \\
%          Adversary side-information resource & limited & unlimited \\
%          Privacy-loss guarantee & on-average & worst-case\\\hline
%     \end{tabular}}
%     \caption{\textcolor{red}{Comparison of threat models}}
%     \label{tab:threat_model}
% \end{table}
Our adversary has the goal of gathering private information that may be revealed by users operating in the mobile data network who are sending data reports to the service provider. The adversary may use this information for purposes not in the users' interest, %such as for \textcolor{red}{targeted marketing}, 
or even to aid criminal attacks such as identity theft. To accomplish his/her goal, the adversary will seek to obtain access to as many user feature reports as possible, consisting of $(u,x_1,x_2,x_3,...)$. Since the primary information sought by the adversary may not be explicitly present in the reports, e.g., if the reports are intentionally obfuscated, the adversary will perform inference attacks to estimate the private user information they desire. The nature of the threat may have some variation dependent on the specific mobile data application and the capabilities of the adversary. With this in mind, we consider the following properties as part of the definition of the threat model:
\begin{itemize}
    \item Whether the adversary can access individual user reports directly, or whether their access is limited to the aggregated reports of all users,
    \item whether the adversary should be assumed to have bounded computational resources,
    \item whether the adversary has access to relevant side information, and
    \item whether users are primarily concerned with potential exposure of private information from their reports on average or in the worst-case.
\end{itemize}
Side information is any additional information that may be available to an adversary that could be used to supplement the information collected from the user reports to increase the efficacy of an inference attack. This could include public databases from organizations like the US Census Bureau or the Department of Transportation which allow an adversary to associate data features, e.g., addresses with names. 
%Since it is intractable to comprehensively model access to possible forms of side information, we allow the adversary to train over obfuscated datasets for which it has access to unobfuscated data, see Section X for more details.

\noindent \textbf{Typical mobile network data threat model:} For most mobile network data applications and users, we apply the following threat model:
\begin{itemize}
    \item The adversary can access individual user reports directly,
    \item the adversary's computational resources are bounded,
    \item the adversary has limited access to side information, and
    \item users are primarily concerned with privacy exposure on average.
\end{itemize}
We consider that many users are likely to have reservations about providing private data to a service provider, either because they do not trust the provider to adequately protect their data or they believe the service provider will themselves use the data in ways that do not align with the user's interests. For this reason, we assume a threat model where users must be able to protect their private information at the local level, e.g., at the user device. We also recognize that some users likely will not have such reservations, and thus a minority of users can be incentivized, e.g., through discounts, to trust a data aggregator with their data, allowing for the possibility of training or tuning privacy schemes based on real user data. Adversary computational resources are assumed to be bounded, recognizing that other methods outside the scope of the data network could be employed to reliably obtain the same private information if an adversary is assumed to have limitless resources. Adversary access to side information is assumed to be limited for the same reason. Finally, we assume users will typically be concerned with the exposure of their private information on average. For most mobile data applications, a user will likely operate with the network over a long period of time and will generate many feature reports as a result. Further, exposure of the private data of any one report will typically pose a much lower risk than exposure through the aggregation of many reports over a period of time. Thus, protecting against an adversary attack on any single report under worst-case conditions is unnecessary for typical applications.
%, and may lead to worse average privacy outcomes by overly constraining the solution space. 

\noindent \textbf{Worst-case mobile network data threat model:} Due to the wide variety of potential mobile network data applications and possible user privacy concerns, we acknowledge there may be some use cases where a worst-case threat model is appropriate. To account for this, we also treat such a model in our analysis. This adversary can access individual reports directly, but in contrast to the typical threat model  
%and all users must be protected, i.e., no users can be incentivized to share their private information with the service provider.  -- see lilly's comment
we assume the adversary has unbounded computational resources and unlimited access to side information. Also, users are concerned with exposure of any single feature report, and their private information in each report must be protected from exposure under worst-case conditions.

%Our adversary, described below, has access to the obfuscated dataset which is, at minimum, stripped of user ID. The adversary attempts to learn private features from this dataset through a combination of classification and estimation. Because the adversary learns patterns which associate public and private features, it is not sufficient to only obfuscate private features. The obfuscation functions considered in this work are applied to full measurements, i.e. noise may be added to all features.

%\textcolor{blue}{Access to individual user reports will necessitate the application of obfuscation schemes at the local user level. This may occur if there does not exist a trustworthy third party to aggregate user data. If an adversary's computational resources and access to side information cannot be assumed to be bounded, then explicitly modeling an adversary is not feasible, and we must rely on metrics for the information exposed by the privatizer to provide a relative measure of how any given adversary may perform against particular privatizers. We discuss this more in Section \ref{metrics}. }

\renewcommand{\arraystretch}{1.3}
\begin{table*}[h]
\centering
\footnotesize
\begin{threeparttable}
\begin{tabular}{p{0.08in}<{\centering}|p{1.8in}<{\centering}|p{1.35in}<{\centering}|p{1.4in}<{\centering}|p{1.4in}<{\centering}|p{0.05in}<{\centering}}
\cline{2-5}
    & 
    %\backslashbox{Properties}{Privatizers} 
    %Privatizers
    & \emph{LDP} & \emph{GAP} & \emph{IT}  & \\\cline{1-5}
\multicolumn{1}{|c|}{\multirow{6}{*}{\rotatebox[origin=c]{90}{Threat model}}} 
& \emph{Adversary computational resources}             
& Unlimited             
& Limited    
& Unlimited    
&         \\ \cline{2-5}
\multicolumn{1}{|c|}{} 
& \emph{Adversary side-info access}                  
& Unlimited             
& Limited    
& Unlimited    
&        \\ \cline{2-5}
\multicolumn{1}{|c|}{}                              
& \emph{Type of privacy-loss guarantee}                         
& Worst-case            
& On-average 
& On-average 
&            \\ \cline{2-6}
\multicolumn{1}{|c|}{}                              
& \emph{Provable adversary privacy protection}      
& Against any adversary 
& Against trained adversary
%Against $P$ (see Eq. (\ref{eq:com_privacy}))
%adversary loss
& Against any adversary 
%Against $I$ (see Eq. (\ref{eq:mi1}))
%mutual info
& \multicolumn{1}{c|}{\multirow{4}{*}{\rotatebox[origin=c]{-90}{Context}}} \\\cline{2-5}
\multicolumn{1}{|c|}{}                              
& \emph{Privatizer access to data for training}              
& Not necessary but helpful*
& Yes        
& No         
& \multicolumn{1}{c|}{}\\ \cline{2-5}
\multicolumn{1}{|c|}{}                              
& \emph{Privatizer access to data distribution}
& Not necessary but helpful*
& No         
& Yes        
& \multicolumn{1}{c|}{}\\ \cline{1-5}
& \emph{Utility protection type}     
& None/Some*
& Maximize utility
%Against $U$ (see Eq. (\ref{eq:com_utility}))
%distortion
& Lower bound on utility
%Against $U$ (see Eq. (\ref{eq:com_utility}))
%distortion
& \multicolumn{1}{c|}{}\\\cline{2-6} 
\end{tabular}
\begin{tablenotes}\footnotesize
\item[*] As discussed in detail in Section \ref{sec:dp}, LDP requires clipping. While clipping can be done in a manner which is agnostic to the data \cite{thakkar2019differentially,mcmahan2018general}, this may result in large utility loss. As a result, clipping is usually performed using information about the data to ensure the added noise is calibrated with the range of data values, see Eq. (\ref{eq:dp4}).
\end{tablenotes}
\end{threeparttable}
\caption{Context used by privatizers (last 4 rows) and properties of threat models (first 6 rows).}
\label{tab:comp_privatizer}
\end{table*}
\renewcommand{\arraystretch}{1}

\subsection{Data Obfuscation and Privatizers}
\label{subsec:pri}
To protect against the adversary threat, privacy can be preserved through obfuscation of the feature data provided by individual users before being released to the service provider. At a minimum, the feature set is stripped of user ID. Remaining features are then obfuscated according to the selected privatization scheme, or "privatizer" for short. This is needed because the adversary may learn patterns in the data which associate public and private features, thus it is not sufficient to only obfuscate private features.
%As a result, all privatizers considered in this work have the capability of altering all features in the user report.

The privatizer will produce an obfuscated measurement feature report $(u,x_1,x_2,x_3,\ldots) \rightarrow (y_1,y_2,y_3,\ldots),$ with $y_i$ denoting the obfuscated version of $x_i$, where the mapping depends on the design of the privatizer. We will consider several privatizers, described fully in Section \ref{sec:privatizer}. Some privatizers leverage actual user data in their design. We assume such data is collected either through opt-in surveys and service provider incentives, or else collected by the provider through other means such as wardriving. In our analysis, we use 70\% of our available dataset for training our adversary (see Section \ref{subsec:adversary} for more details) as well as for training, fitting models, and/or choosing parameters of the privatizers (see Section \ref{sec:privatizer} for more details). The remainder of the dataset will be used to test our privatizers against the adversary. 

\subsection{Context}
\label{sec:context}

User data is a type of application-specific context, and different privatizers may use the actual data, data distributions, or merely data moments like mean and variance. There are other types of application-specific context, e.g. privacy and utility metrics of interest, which privatizers may optimize over. 
Since mobile service providers know what they want to use the data for, and may ask their clients about privacy concerns, such metrics may indeed be available to be used in the design of privatizers.

Using context has implications to the threat model. For example, optimizing over a particular privacy metric guarantees protection against this privacy metric but not against any function of the data.
As another example, if a privatizer optimizes its design under a known data distribution, or is trained under a given dataset, its performance is not guaranteed under different distributions and datasets. 

Using context may also offer utility guarantees since optimizing over, or putting a constraint on a utility metric, restricts the privatizer from making obfuscation decisions that reduce utility below acceptable levels. %sizably. 

Table \ref{tab:comp_privatizer} compares different privatizers with respect to how much context they use and which threat model properties they can protect against. 
LDP offers stronger privacy protection than the rest as it provides worst-case privacy guarantees against any adversary with potentially unlimited resources and side information access. 
However, it does not have a formal mechanism to guarantee a minimum level of utility. 
In contrast, GAP and IT are aware of application specific utility metrics which they include in their optimization setups, and thus provide utility guarantees. 
The GAP privatizer in particular optimizes a multi objective function which considers both privacy and utility. That said, it is optimized and can offer formal guarantees only against the particular adversary in its training loop. 
These fundamental distinctions among the different obfuscation approaches are discussed in more detail in Section \ref{sec:privatizer} and their implications to the privacy-utility tradeoff are presented and discussed in detail in Section \ref{sec:result}.
%While this table refers to metrics defined in Section \ref{sec:metric} and privatizers defined in Section \ref{sec:privatizer}, we present it here as it is major component of our system model.

\subsection{Adversary Model}
\label{subsec:adversary}

Depending on whether users upload measurements to the server one at a time or in batches, the adversary may or may not know whether a sequence of measurements originated from a single user or multiple users.
Consider first the scenario where 
each user uploads one obfuscated measurement each time. Given that the user ID of each obfuscated measurement is unknown, the adversary takes as input one measurement from the obfuscated dataset $(y_1,y_2,y_3,\ldots)$ and predicts the user ID and true location (the unobfuscated latitude and longitude) from which the measurement originated $(\hat{u},\hat{x}_1,\hat{x}_2,,\hat{x}_3\ldots)$.
Now consider the scenario where, for the sake of reduced system complexity, each user uploads a sequence of obfuscated measurements each time\footnote{In practice, the service provider can require users to upload their data weekly or monthly.}. While the user ID of each obfuscated measurement in the database is unknown, the adversary knows that measurements in the same batch belong to the same user, and can take advantage of correlations across measurements to improve estimation. In this case the adversary takes as input a measurement sequence $\{(y_{1i},y_{2i},y_{3i},\ldots)\}_{i=1}^{i=L}$ from a single user and predicts a single user ID $\hat u$ and the true locations $\{(x_{1i},x_{2i})\}_{i=1}^{i=L}$ ($i$ denotes the $i^{th}$ measurement in this sequence, and $L$ is the sequence length).
In Section \ref{sec:result} we 
investigate the performance under both scenarios, see Section \ref{subsec:sequence} for a direct comparison between the two.
%In implementation, the adversary operates on random batches of measurements, but predictions are made independently per measurement. In Section \ref{sequencesection} we discuss an adversary which makes joint predictions for a series of measurements.

\begin{figure}{
    \centering{
    \includegraphics[width=\columnwidth]{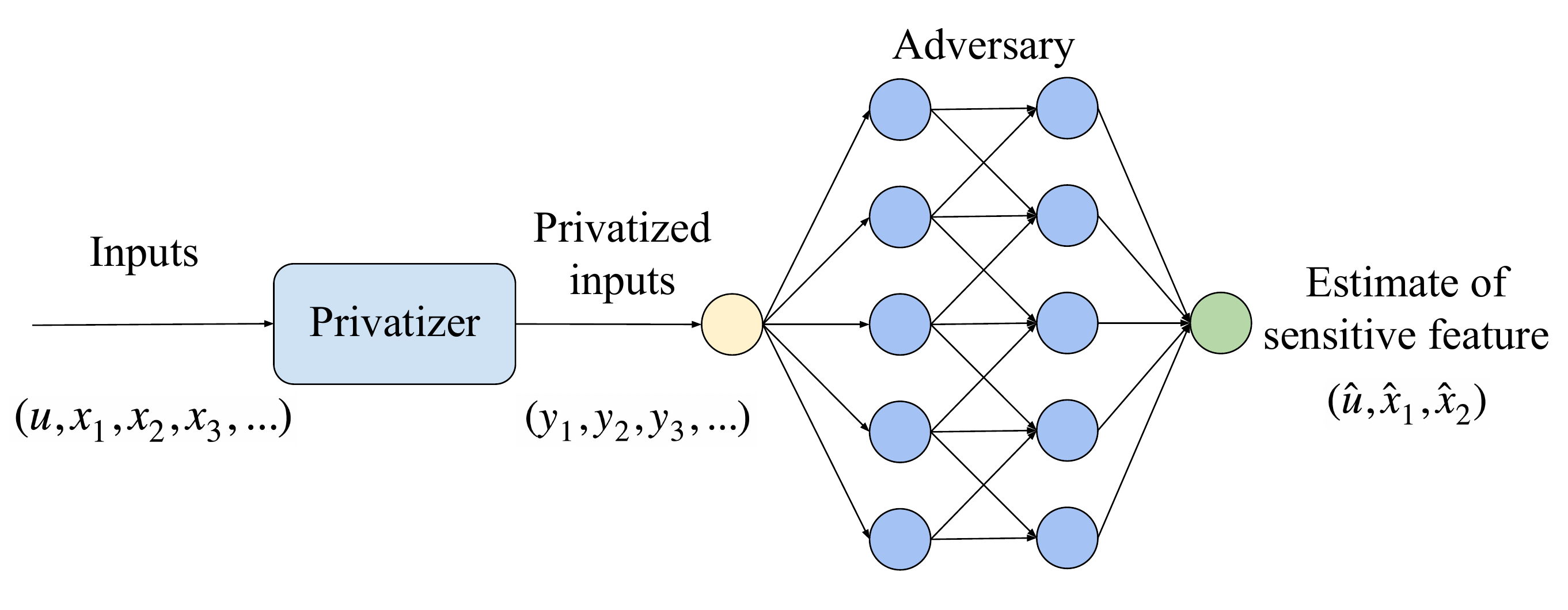}}
    \caption{A diagram of adversary implementation.}
    \label{fig:adversarynn}}
\end{figure}

The adversary estimation is a mapping from $(y_1,y_2,y_3,\ldots)$ to  
$(\hat{u},\hat{x}_1,\hat{x}_2)$ and one may use a number of approaches to perform that mapping. In theory, one may discretize the continuous $x_i$'s and $y_i$'s and use empirical conditional probabilities and maximum likelihood estimation, but in practice the state space would explode. Given the availability of real world datasets, learning is a better choice.
We experimented with linear and non-linear models for used ID estimation, and chose a deep neural network (DNN) to model our adversary (see \F\ref{fig:adversarynn}), given the effectiveness of DNNs in approximating non-linear functions.

Specifically, 
our adversary is modeled as a fully-connected DNN containing two hidden layers with 256 neurons each. Between layers we employ Rectified Linear Unit (ReLU) activations, and our optimization relies on Adaptive Moment Estimate (Adam) stochastic gradient descent with a learning rate of 0.001. These values were empirically selected to maximize the adversary's performance when given the unobfuscated data as input.
% Note: if it is not a discrete mapping, we can not use classical discrete estimation method like MLE
% We could use linear model and non-linear model, and but the non-linear estimation is better. So we choose the non-linear model, which will be more conservative, since it is a stronger adversary.

Assume that the input measurement contains $m$ features and there are $k$ users ($m$ and $k$ depend on three datasets described in Section \ref{subsec:userdata}). Then each input batch has $n$ measurements containing the $m$ features. The output of the adversary neural net is a $n\times (k+2)$ matrix representing estimates of user ID and location ($n=1024$ in our experiments). Each row in this matrix contains the likelihood that this measurement belongs to different users, and the estimated latitude and longitude of the original measurement. The loss function used to train the adversary is a weighted sum of the categorical cross entropy loss of the user ID estimate vector and the euclidean distance between the actual location and the location estimate. The user ID estimate error, location estimate error, and adversary loss functions are defined in Section \ref{sec:metric}. 

We provide our adversary 70\% of the obfuscated dataset to train on, for which it has access to the unobfuscated user IDs and locations,
%The adversary trains for five epochs (limited to prevent overfitting), 
and test it on the remaining 30\% of the data. 
Providing the adversary such a high portion of the data for training makes our privacy results conservative. In our threat model we have assumed some access to side information but comprehensively modeling access to possible forms of side information is intractable. %The adversary's access to 70\% of the dataset with obfuscated and true user ID and location labels serves as an approximation of access to side information in the following sense:
%Supervised learning on such a training set can be thought of as a conservative bound on an adversary's ability to learn the system by observing it for an extended period of time leveraging side information such as known user locations at certain times, or by running it's own devices on the network to establish ground truth, or by simulating published privatizers as they may be revealed by the service provider to help convince users regarding their ability to preserve privacy.
The adversary’s access to 70\% of the dataset with obfuscated and true user ID and location labels serves as an approximation of some form of side information. 
%Access to the data set can be thought of as the adversary observing the system for an extended period of time leveraging side information. 
Side information may include known user locations at certain times, or inputs from the adversary's own devices on the network to establish ground truth. The training set could also correspond to the adversary simulating published privatizers which may be revealed by the service provider to help convince users regarding their ability to preserve privacy.

\subsection{Signal Map Model}
\label{subsec:service}
The service provider trains an RSS predictor based on the the aggregated user data such that it can generate accurate signal maps. Specifically, the model input features include (obfuscated) latitude, longitude and other features (i.e. %$(x_1, x_2, x_{j,\mbox{} j>3})$
$(x_1, x_2, x_j,j>3)$), and the model output is the RSS value $x_3$ in dBm.

There is a long line of research on RSS predictor models, see, for example \cite{erceg1999empirically,han2013double,alimpertis2019city}. We first consider 
a simple path loss model \cite{molisch2012wireless} but find its accuracy to be underwhelming. We also consider a linear and a neural network model and find that both have good comparable accuracy yet the former is easier to work with. Notably, its parameters can be estimated in one step which allows us to calculate application-specific utility metrics more efficiently (see Section \ref{subsec:utility}). We thus select a linear RSS prediction model. Specifically, we use the following model:
\begin{equation}
    x_3 = a_0 +  \sum_{j=1, j\neq 3}^{j=m}a_{j-1}x_j,
\end{equation}
where $k$ is the total number of features in a measurement and $\alpha=[a_0,...,a_{m-1}]^T$ is the parameter vector. Given a set of $n$ measurements $X=[x_{ji}]$ where $j=1,...,m$ and $i=1,...,n$ ($m$ is the number of features), the parameter vector of the RSS prediction model can be estimated via linear regression as follows:
\begin{equation}
    \alpha_X=(X_{-3}^TX_{-3})^{-1}X_{-3}^TX_3,
    \label{eq:alpha}
\end{equation}
where $X_3$ is the third column of $X$ and $X_{-3}$ is the remaining columns of $X$ without the third column. Similarly, given a set of $n$ obfuscated measurements $Y=[y_{ji}]$ where $j=1,...,m$ and $i=1,...,n$, the parameter vector of RSS prediction model can be estimated as $\alpha_Y$.

% Note: we choos elinear RSS linear prediction because it has better accuracy performance, we have tried a path-loss model based prediction. We consider a simpler path model, but the RMSE is higher. We also consider a more complex neural network.

\section{Definition of Metrics} \label{metrics}
\label{sec:metric}
In this section we define the metrics used to evaluate privacy and utility.

\subsection{Privacy}
%\textcolor{blue}{For the bounded adversary threat model,}
Let $n$ denote the number of measurements per batch. $u=[u_i]$, $i=1 \ldots n$ represent the user ID of each collected measurement and $\hat{u}=[\hat{u}_i]$ is the adversary's estimate of $u$.
The adversary computes a probability distribution over the space of possible user
IDs and selects for each measurement the user ID estimate with the maximum likelihood. 
We define the adversary estimate accuracy as the fraction of correct user ID estimates, that is,
\begin{equation*}
%\label{acc}
    acc(\hat{u},u) = \frac{1}{n} \sum_{i=1}^{n} 1_{\hat{u}_i = u_i},
\end{equation*}
where the indicator function
$1_{\hat{u}_i = u_i}$ is equal to 1 if the estimate is correct and 0 otherwise.
%The factor of 100 converts this to a percentage.
Since high values of accuracy correspond to low values of privacy, we define the first privacy metric as 
\begin{equation}
    P_1(\hat u,u) = 1 - acc(\hat u,u).
\end{equation}

$\hat{x}_1$ and $\hat{x}_2$ are the adversary's estimates of the true latitude $x_1$ and longitude $x_2$. While $\hat{u}$ represents a probability distribution, $\hat{x}_1$ and $\hat{x}_2$ specify an exact location. 
Our second privacy metric is the Euclidean distance between the true location and the adversary's estimate
averaged over the batch, defined by
\begin{equation}\label{eq:p2}
    P_2(\hat{x}_1,\hat{x}_2,x_1,x_2) = \frac{1}{n} \sum_{i=1}^{n}{ \sqrt{ (\hat{x}_{1i}-x_{1i})^2 + (\hat{x}_{2i}-x_{2i})^2}},
\end{equation}
where the subscript $i=1...n$ corresponds to the $i^{th}$ measurement in the batch of size $n$.
This metric defines how well the adversary is able to recover the original user location. High values of adversary location error correspond to high privacy.

% \mattdraft{DID WE WANT TO REMOVE THE DEFINITIONS OF THE COMPOSITE METRICS HERE AND JUST INTRODUCE THEM IN THE RESULTS? }
Since both user IDs and locations are considered as private and sensitive information in our application, we further define the following composite privacy metric:
\begin{equation}
\label{eq:com_privacy}
    P(\hat{x}_1,\hat{x}_2,\hat{u},x_1,x_2,u) = v_1P_1(\hat{u}, u) + v_2P_2(\hat{x}_1,\hat{x}_2,x_1,x_2),
\end{equation}
where $v_1$ and $v_2$ are parameters controlling the weights of the two aforementioned privacy metrics.
$P_1, P_2$ and $P$ are the privacy metrics we use throughout the paper to compare the performance of different privatizers.
%\textcolor{red}{Lilly you have to be mathematically clear with notation on a journal paper so add a sentence along the lines of "where subscript $i$, $i=1 \ldots n$, refers to the corresponding measurement" or something along these lines.}

% Both of the privacy metrics are important in our application.

\subsubsection{Additional privacy metrics}
The composite privacy metric defined above is not differentiable because $P_1$ is not differentiable. This is a problem for adversary training. To handle this we use the cross entropy loss of the user ID estimate
\begin{equation}\label{eq:p1}
    % P_1^{ce}(\hat{u}, u) = \frac{1}{n}\sum_{i=1}^{n}{\sum_{j=1}^{l}{1_{\underset{j}{\mathrm{argmax}}(\hat{u}_{ij}) = u_i}(-\log{\hat{u}_{ij}})}} \\
    P_1^{ce}(\hat{u}, u) = -\frac{1}{n}\sum_{i=1}^{n}
    \log{p_i},
\end{equation}
where $p_i$ is the estimated likelihood of user ID $u_i$ for measurement $i$,
and define the loss function of the adversary as
\begin{equation}\label{eq:La}
    L_a(\hat{x}_1,\hat{x}_2,\hat{u},x_1,x_2,u) = v_1 P_1^{ce}(\hat{u}, u) + v_2 P_2(\hat{x}_1,\hat{x}_2,x_1,x_2),
\end{equation}
which is used in the training of the adversary and of the GAP neural networks (the GAP privatizer and adversary used in the iterative training, see Section \ref{sec:gap}). 
%where $v_1^a$ and $v_2^a$ are parameters controlling the weights of two loss items like before.
%We experimented with weighting the two components of the loss function, 
% however empirically determined that the adversary trained with the unweighted loss function performs equally well.
% however empirically determined that the adversary trained with the unweighted loss function offers a good balance of performance with respect to both estimation of user IDs and location.

% For illustrative purposes, and because it is an intuitive metric closely related to $P_1$, we consider as well the adversary estimate accuracy, defined by
% \begin{equation}\label{acc}
%     acc(\hat{u},u) = \frac{100}{n} \sum_{i=1}^{n}\sum_{j=1}^{l}{1_{\underset{j}{\mathrm{argmax}}(\hat{u}_{ij}) = u_i}},
% \end{equation}
% where the average is taken over the batch of size $n$. The factor of 100 converts this to a percentage.

Our IT privacy approach, see Section \ref{sec:infotheory}, is motivated by the use 
of mutual information as a measure a privacy. 
The mutual information between two random variables $X$ (e.g., our input) and $Y$ (e.g., the obfuscated data) quantifies how much information about one random variable is obtained through observing the other. It is given by
\begin{equation}\label{eq:mi1}
    I(X;Y) = \sum_{y\in\mathcal{Y}}{\sum_{x\in\mathcal{X}}{p_{X,Y}(x,y)\log\Big(\frac{p_{X,Y}(x,y)}{p_X(x)p_Y(y)}\Big)}},
\end{equation}
where $p_{X,Y}$ is the joint probability mass function and $p_X, p_Y$ are the marginal probability mass functions.
%Considering $X$ and $Y$ as random variables describing our input and obfuscated data respectively, minimizing $I(X;Y)$ can be interpreted as providing privacy on an average sense.

Last, the privacy metrics defined above are well suited for the typical threat model discussed in Section \ref{subsec:threatmodel}. However, for the worst-case threat model involving adversaries with unbounded computational resources and auxiliary information where users seek protection of any single report (see Section \ref{subsec:threatmodel}) we resort to differential privacy (DP) \cite{dpresults}. 
%Intuitively, differential privacy ensures that the addition or removal of any database item does not (substantially) affect the outcome of any analysis \cite{dpresults}. Formally,
Specifically,
let $\mathit{K}$ be a randomized function applied to the input dataset. $\mathit{K}$ gives $\epsilon$,$\delta$-differential privacy if for all datasets $\mathit{D_1}$ and $\mathit{D_2}$ which differ in at most one element and $\forall S \in \textit{range}(\mathit{K})$,
\begin{equation}\label{eq:dp1}
    Pr[K(D_1) \in S] \leq e^{\epsilon}Pr[K(D_2) \in S] + \delta,
\end{equation}
where the probability is taken over the randomness in $\mathit{K}$. $\epsilon$ and $\delta$ bound the difference between the output of $\mathit{K}$ on $\mathit{D_1}$ and $\mathit{D_2}$ thus making it hard to guess the input ($\mathit{D_1}$ versus $\mathit{D_2}$) by observing the output. 
DP is a strong guarantee, since it doesn't make any assumptions about the computation power and auxiliary information available to the adversary, and $\epsilon$ and $\delta$ serve as metrics for privacy, see \cite{dp} for more details.

%***********************
\subsection{Utility}
\label{subsec:utility}
Let $m$ be the number of features at each measurement excluding the user ID which is stripped from the input. The output of the privatizer $y=[y_j]$, $j=1\ldots m$ is the obfuscated data, e.g. $y_1$ and $y_2$ denote obfuscated latitude and longitude, respectively. Our utility metrics quantify the difference between the input data $x=[x_j]$ and the obfuscated data $y=[y_j]$. Recall that we consider $n$ measurements per batch thus $x_j$ and $y_j$ are vectors of size $n$ with elements $x_{ji}$ and $y_{ji}$, $i=1\ldots n$, respectively. 
We consider several utility metrics motivated by real-world applications of crowdsourced network data.

The first metric quantifies the overall distortion of the dataset, considering all $m$ features, by the L2 norm distance between input and obfuscated data, averaged over all $n$ batch measurements:
\begin{equation}\label{eq:u2}
    % U_1(x,y) = -\frac{1}{n} \sum_{i=1}^{n}{ \vert\vert x_i-y_i \vert\vert_2},
    U_1(x,y) = -\frac{1}{n} \sum_{i=1}^{n}{ \sqrt{\sum_{j=1}^{m}(y_{ji}-x_{ji})^2} }.
\end{equation}
%the subscript $i=1...n$ denotes the $i^{th}$ measurement in the batch, the subscript $j=1...m$ denotes the $j^{th}$ feature of , such that $y_i$ corresponds to $x_i$ after obfuscation. 
Intuitively, high values of distortion correspond to low utility thus the minus sign in front of distortion in Eq. (\ref {eq:u2}).

The second utility metric is related to the RSS prediction model described in Section \ref{subsec:service}. Recall that the goal of service provider is to estimate an accurate RSS prediction model based on the aggregated user data. However, with obfuscated user data, the estimated parameters of RSS prediction model differs from those estimated by unobfuscated user data (i.e. the estimated parameter vector changes from $\alpha_X$ to $\alpha_Y$, see \Eq(\ref{eq:alpha})). To minimize the difference between them, we define our second utility function as the opposite of L1-norm distance between $\alpha_X$ and $\alpha_Y$ as follows:
\begin{equation}\label{eq:u1}
    U_2(x,y) = -\vert\vert{ \alpha_X - \alpha_Y}\vert\vert_1.
\end{equation}
where $\alpha_X$ represents the RSS prediction model parameters estimated by unobfuscated user data (i.e. the privatizer's input) and $\alpha_Y$ represents the RSS prediction model parameters estimated by obfuscated user data (i.e. the privatizer's output). We refer to $\vert\vert{ \alpha_X - \alpha_Y}\vert\vert_1$ as the generated map error, where higher values of map error corresponds to lower utility. While many metrics could be used to measure the distance between $\alpha_X$ and $\alpha_Y$, comparing the fitted parameters over this bounded space provides a simple, effective loss function. 
Note that this map error does not capture how well a RSS prediction model generated by the obfuscated data could be used to predict RSS values at a new location, but rather captures the ``distance" between maps generated before and after obfuscation. 
% we envision that a composite metric will be practical to the service provider.

Envisioning that the service provider may care for more than a single application-specific utility metric like $U_2$ in practice, we further define a composite utility metric $U(x,y)$ as
\begin{equation}
\label{eq:com_utility}
\begin{aligned}
&U(x,y) = w_1U_1(x,y) + w_2U_2(x,y),
\end{aligned}
\end{equation}
where $w_1$ and $w_2$ are parameters adjusting the weights of each utility metric.
%We will revisit these privacy and utility metrics when analyzing the performance of each of our four privatizers.

%\begin{equation}
%    C_{jk}(x) = \sum_{i=1}^{n}{1_{(x_{i1},x_{i2}) \in c_{jk}}},
%\end{equation}
%where $1_{(x_{i1},x_{i2}) \in c_{jk}}$ is an indicator function which is equal to 1 if the $i^{th}$ point in $x_1, x_2$ exists in the $j^{th}$ row and $k^{th}$ column of the grid, and the sum is taken over the $n$ points in the batch. 
\section{Privatizers}
\label{sec:privatizer}
In this section we introduce in detail each of the four privatizers, which represent different types of obfuscation schemes. Specifically, we first select a Gaussian noise-adding privatizer for its simplicity and as a benchmark. We then select a locally differentially private privetizer  (LDP) motivated by the well known strengths of Differential Privacy. We then select a privatizer based on GANs (referred to as the GAP privetizer), given the recent interest on how adversarial learning may be used to train privatizers by positioning them against adversities. Last, we select the so-called IT privatizer since it is a good representative of obfuscation schemes which use mutual information as a privacy metric and optimization to optimally design obfuscation.

\subsection{Gaussian Noise-Adding Privacy}

\begin{figure*}
\centering
\begin{subfigure}{.495\textwidth}
    \includegraphics[width=1\linewidth]{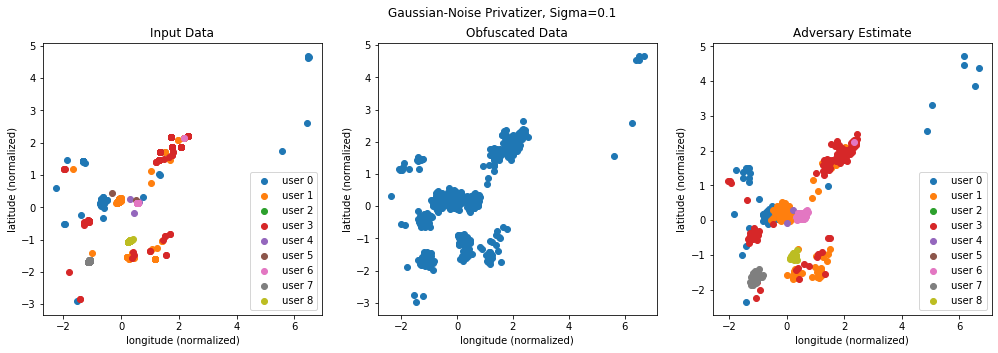}
    \caption{Input Data, Obfuscated Data, Adversary Estimate}
    \label{fig:lownoise}
\end{subfigure}
\begin{subfigure}{.495\textwidth}
    \includegraphics[width=1\linewidth]{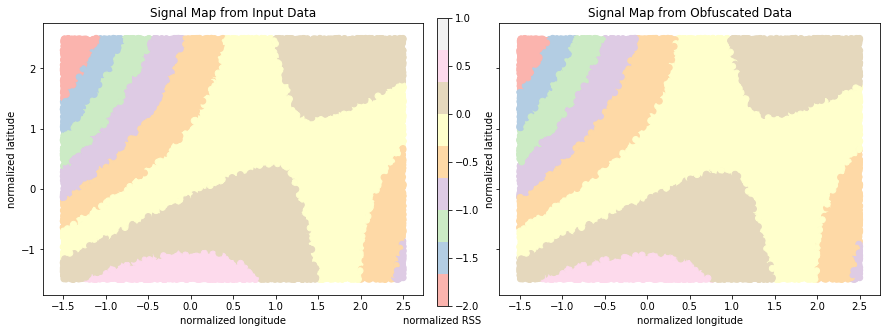}
    \caption{Signal Maps Before and After Obfuscation}
    \label{fig:lownoisemap}
\end{subfigure}
\caption{Noise privatizer when $\sigma$=0.1.}
\label{fig:noise_privatizer}
\end{figure*}

\begin{figure*}
\centering
\begin{subfigure}{.495\textwidth}
    \includegraphics[width=1\linewidth]{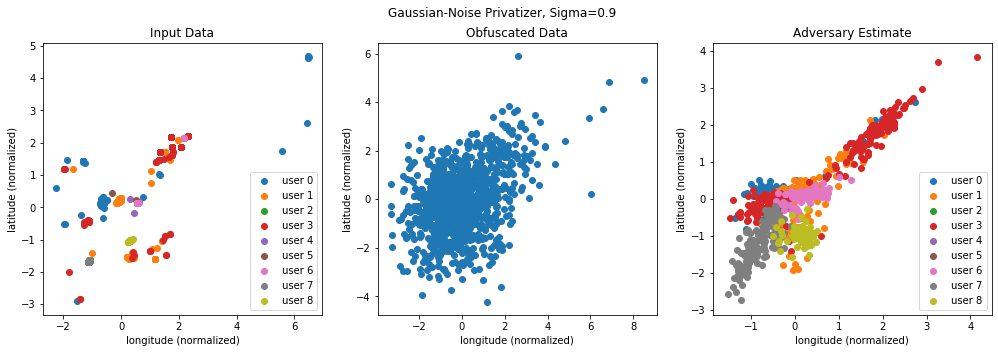}
    \caption{Input Data, Obfuscated Data, Adversary Estimate}
    \label{fig:highnoise}
\end{subfigure}
\begin{subfigure}{.495\textwidth}
    \includegraphics[width=1\linewidth]{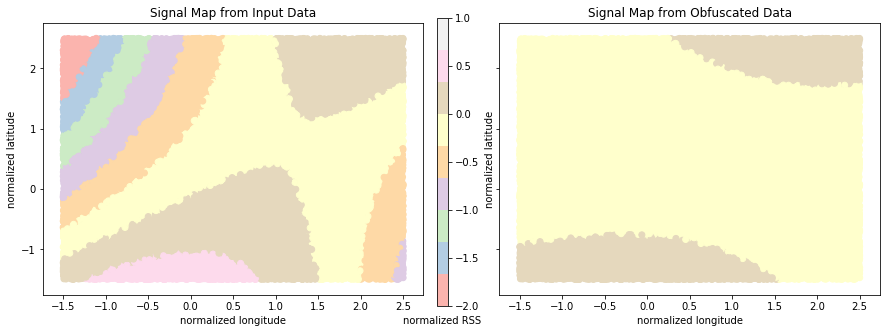}
    \caption{Signal Maps Before and After Obfuscation}
    \label{fig:highnoisemap}
\end{subfigure}
\caption{Noise privatizer when $\sigma$=0.9.}
\label{fig:noise_privatizer}
\end{figure*}

Our Gaussian noise-adding privatizer (Noise privatizer) takes the simplest approach to data obfuscation. For each input batch of size $n \times m$, where $n$ is the number of points and $m$ is the number of features, we add an $n \times m$ matrix of Guassian noise. Each element in this noise matrix is normally distributed with a mean of 0 and a standard deviation of $\sigma$. Since the data is also normalized such that each feature has a mean value of zero with a standard deviation of 1, values of $\sigma$ close to 1 add a significant amount of noise and we choose to vary $\sigma$ between 0 and 1. \F\ref{fig:lownoise} provides a visualization of what the normalized input data, obfuscated data, and adversary's estimate look like side by side using the Noise privatizer for a low value of $\sigma$. \F\ref{fig:lownoisemap} shows the signal maps generated before and after obfuscation. This shows that even in the presence of obfuscation, we can generate representative signal maps with the obfuscated data. Figures \ref{fig:highnoise} and \ref{fig:highnoisemap} show the same plots for a high value of $\sigma$. Note that while the privacy is improved, i.e. the adversary estimate is further from the input data, the signal maps differ significantly.

For reference, a privatizer which releases a completely random dataset (from a normal distribution with variance of 1.0) regardless of input data would observe the errors shown in Table \ref{table:worstcaseutility}. 
\renewcommand{\arraystretch}{1.3}
\begin{table}[h]
\caption{Utility Reference Values}
\centering
\footnotesize
\begin{tabular}{c c c}
\hline\hline
Metric & No Obfuscation & Random Data\\ [0.5ex]
\hline
Distortion ($-U_1$) & 0 & 5.74 \\
Generated Map Error ($-U_2$) & 0 & 2.41 \\
%Geographic Density Error & 0 & 0.90 \\
\hline
\end{tabular}
\label{table:worstcaseutility}
\end{table}
\renewcommand{\arraystretch}{1}

% \subsubsection{Methodology}


\subsection{Local Differential Privacy}
\label{sec:dp}

% \subsubsection{Methodology}
We implement two local DP (LDP) privatizers which provide mathematical guarantees for privacy (see Eq. (\ref{eq:dp1})) under the worst-case threat model discussed in Section \ref{subsec:threatmodel}. 

The first approach is the Gaussian Mechanism with parameters $\epsilon$ and $\delta$, which we refer to as GLDP \cite{dp}. This mechanism adds zero-mean Guassian noise with variance $b$ to each feature. This variance is defined by
\begin{equation}\label{eq:dp2}
    b = \frac{\Delta^2}{\epsilon^2}2\ln(\frac{1.25}{\delta}),
\end{equation}
where $\Delta$ is the L2-sensitivity of a function $f$ given by
\begin{equation}\label{eq:dp3}
    \Delta = \max_{D_1,D_2\in D}{\vert\vert f(D_1) - f(D_2)\vert\vert_2},
\end{equation}
where $D_1$ and $D_2$ are subsets of the dataset $D$ differing only by one element. Generally, in the local DP model, one can think of $D_1$ and $D_2$ as datasets of size 1 (i.e. one data point) and $f$ as an identity function. Therefore the sensitivity becomes the greatest L2-distance between any two data points.
In practice, we use an analytically calibrated Gaussian Mechanism which is shown to extend better to the very high privacy regime $(\epsilon \longrightarrow 0)$ and the low privacy regime $(\epsilon \longrightarrow \infty)$, see Algorithm 1 in \cite{analytical_gaussian} for the exact calculation for the variance of the added noise $b$.

The second approach is the Truncated Laplacian Mechanism with parameters $\epsilon$ and $\delta$, which we refer to as LLDP, recently proposed in \cite{geng2018privacy}. This mechanism adds noise satisfying the truncated Laplacian distribution, with probability density function
\begin{equation}
\label{eq:tldp1}
f(x)=
\begin{cases}
    Be^{-\frac{|x|}{\lambda}},\quad for \quad x\in[-A,A]\\
    0,\quad otherwise
\end{cases}
\end{equation}
where
$$\lambda = \frac{\Delta}{\epsilon},\quad A=\frac{\Delta}{\epsilon}\log(1+\frac{e^{\epsilon}-1}{2\delta}), \quad B=\frac{1}{2\lambda(1-e^{-\frac{A}{\lambda}})},$$ and $\Delta$ is defined in Eq. (\ref{eq:dp3}).
%  where $\delta$ is subpolynomially small

For both approaches, we follow standard practice and use $\delta = 0.00001$ ($\delta$ should be much smaller than $1/n$ \cite{dptutorial}) and $\epsilon$ between 1 and 10 (larger 
$\epsilon$ values yield a very loose bound, see Eq. (\ref{eq:dp1})), where low values of epsilon guarantee better privacy.

Moreover, following standard practice again, we clip each data point to have L2-norm $\leq \frac{\Delta}{2}$. Then, by invoking the triangle inequality, we can ensure that sensitivity is no greater than $\Delta$.
Specifically, for both the Gaussian mechanism and Laplacian mechanisms,
we clip each data point according to the following function
\begin{equation}\label{eq:dp4}
    x_{new} = \frac{x}{\vert\vert x \vert\vert_2}\min{(\frac{\Delta}{2}, \vert\vert x \vert\vert_2)}.
\end{equation}
To choose $\frac{\Delta}{2}$, we use the rule of thumb that clipping should occur 5\% of the time. Using the pilot dataset to approximate how much of the data would be clipped for a given value, we choose $\frac{\Delta}{2} = 7.154$ and use this parameter during testing.

\T\ref{tab:ldp_com} compares the GLDP and LLDP privatizers with respect to our privacy and utility metrics on Chania dataset.  We notice that both GLDP and LLDP privatizers yield quite large utility losses. From this table, it is evident that GLDP achieves sizably higher privacy than LLDP w.r.t. $P_1$ and $P_2$, especially for larger $\epsilon$ values. Although GLDP privatizer has larger loss in utility, both GLDP and LLDP privatizers can not offer any utility protection. Hence, we use GLDP with higher privacy in the rest of the paper when comparing LDP with other approaches under the typical threat model, see Section \ref{sec:result}.

Note that while our Noise and LDP privatizers both add normally distributed noise, the key difference between the two is the noise clipping step. Intuitively, this ensures that no two data points are too different. This gives an anonymity to each measurement that is crucial to privacy under a worst-case threat model.
\begin{table}[!h]
\centering
\footnotesize
\caption{Comparison of GLDP and LLDP privatizer on Chania dataset. 
%for all these v1=v2=0.5
}
\renewcommand{\arraystretch}{1.3}
\begin{tabular}{p{0.4in}<{\centering}p{0.5in}<{\centering}p{0.3in}<{\centering}p{0.3in}<{\centering}p{0.3in}<{\centering}p{0.3in}<{\centering}}
\hline\hline
$\epsilon$ & Mechanism & $P_1$ & $P_2$ & $U_1$ & $U_2$\\\cline{1-6}
\multirow{2}{*}{1} & GLDP & 0.68 & 0.94 & 113.76 & 2.96  \\
 & LLDP & 0.68 & 0.94 &	84.60 &	2.95  \\\cline{1-6}
 \multirow{2}{*}{10} & GLDP & 0.63 &	0.91 & 16.46 &	2.39 \\
 & LLDP & 0.49 & 0.69 & 8.50 & 2.49 \\\cline{1-6}
\multirow{2}{*}{100} & GLDP & 0.32 & 0.36  & 4.21 & 2.44 \\
 & LLDP & 0.05 & 0.10 & 0.93 & 2.20 \\\cline{1-6}
 
%  \multirow{2}{*}{$\epsilon=5$} & GLDP & 0.68 & 0.94 & 2.48 & 30.24 \\
%  & LLDP & 0.62 & 0.90 & 2.48 & 16.95 \\\cline{1-6}
\end{tabular}
\label{tab:ldp_com}
\end{table}
\renewcommand{\arraystretch}{1}

\subsection{Generative Adversarial Privacy}
\label{sec:gap}
% \subsubsection{Methodology}
Generative Adversarial Privacy is a data-driven approach to obfuscation which learns a privatization strategy by positioning the privatizer and adversary against each other in a minimax game \cite{gap,cagap}. Our privatizer is a fully-connected feedforward neural network with a similar structure to our adversary. It has two hiddens layers of 256 units each. Between layers we employ Rectified Linear Unit (ReLU) activations, and our optimization relies on Adaptive Moment Estimate (Adam) stochastic gradient descent with a learning rate of 0.001. Our privatizer, which takes an input batch of size $n \times m$, outputs an $n \times m$ batch of obfuscated data, where each measurement has been obfuscated independently. (We treat the case where measurements are grouped into batches and then jointly obfuscated in Section \ref{subsec:sequence}.)

Our privatizer wants to minimize the following loss function
\begin{equation}
\begin{aligned}
&L_p(x,y,\hat{u},\hat{x}_1,\hat{x}_2) = -\rho U(x,y)\\
&-(1-\rho)L_a(\hat{x}_1,\hat{x}_2,\hat{u},x_1,x_2,u),
\label{eq:lp}
\end{aligned}
\end{equation}
where $U$ is the composite utility metric defined in Eq. (\ref{eq:com_utility}) and 
$L_a$ is the adversary loss function defined in Eq. (\ref{eq:La}) which is a differentiable version of the composite privacy metric and depends on the adversary estimate error of the user ID and location. 

Notice that as the adversary's loss decreases (implying less privacy), the privatizer's loss increases.
%In the privatizer loss function, 
$\rho$ quantifies the penalty on utility loss, as opposed to privacy loss. Utility losses have a large effect on the privatizer when $\rho \longrightarrow 1$ and privacy losses have a large affect on the privatizer when $\rho \longrightarrow 0$. 

We take an iterative approach to training the two neural networks. We first train the adversary, specifically, we fix the neural network (NN) weights of the privatizer and perturb the NN weights of the adversary along the negative gradient of $L_a$ for $k$ epochs. We then train the privatizer, that is, we perturb the NN weights of the privatizer along the negative gradient of $L_p$ for $k$ epochs, and so on and so forth. When both have converged, we have found the equilibrium of our minimax game. We then fix the weights of both NNs during testing.

% \subsubsection{Performance}
% Varying $\rho$ characterizes the privacy-utility tradeoff. \F\ref{fig:util_rho} (left y axis) shows the performance of our GAP privatizer with respect to our two privacy metrics, for $k=5$ and convergence defined as $\theta \leq 0.0001$ where $\theta$ is the difference in loss function value during two consecutive training iterations. \F\ref{fig:util_rho} (right y axis) shows the performance of our GAP privatizer with respect to our three utility metrics. 
% Comparing to LDP, GAP can achieve comparable privacy, for low values of $\rho$, for much higher utility. For example, targeting a high privacy regime, for $0\leq \rho \leq 0.4$ the adversary user ID estimate error under GAP is between 50\% and 70\% and the distortion is between 1.4 and 4, whereas for the same level of privacy (adversary user ID error) LDP's distortion ranges between 20 and 70 (at least 15 times higher).
% When comparing GAP to the Noise privatizer, we observe that for comparable utility, GAP achieves better privacy. For example, for a level of distortion of about 2.0 the adversary estimate error against the Noise privatizer is about 15\% down from 50\% against the GAP privatizer, i.e. GAP is almost three times as private for this level of utility.

% In summary, we observe that GAP occupies a more desirable space in the privacy-utility tradeoff space with relatively high privacy and utility (under the typical threat model with a bounded adversary). 
The GAP privatizer incorporates the privacy and utility metrics in its loss function $L_p$ and trains against an adversary with the same loss function $L_a$ as the one used to evaluate privatizers. 
% While the former is a feature, the latter may be perceived as unfair advantage, thus we evaluate its performance against other loss functions too, see Section \ref{sec:different_adversaries}.
While it is advantageous to incorporate specific privacy metrics, for generality we evaluate the GAP privatizer's performance against other loss functions too, see Section \ref{sec:different_adversaries}.

% \begin{figure}{
%     \centering{
%     \includegraphics[width=0.9\columnwidth]{images/gap_chania.eps}}
%     \caption{GAP privatizer: Privacy and utility varying $\rho$ on Chania dataset. $P_1$: user ID estimate error, $P_2$: location estimate error, $-U_1$: distortion, $-U_2$: map error, $-U_3$: location error. }
%     \label{fig:util_rho}}
% \end{figure}

\subsection{Information-theoretic Privacy}
\label{sec:infotheory}

% \subsubsection{Methodology}

For this approach, we consider the privacy-utility tradeoff in an analytically tractable fashion under a formal optimization framework. 

Considering $X \in \mathcal{X}$ and $Y \in \mathcal{Y}$ as random variables describing our input and obfuscated data respectively, our IT privatizer tries to minimize the mutual information $I(X;Y)$, see Eq. (\ref{eq:mi1}), subject to a constraint on utility. The privatizer is specified by the conditional probability distribution $p_{Y|X}$, the probability of releasing $Y$ given input data $X$. Without the utility constraint, the optimal solution is to release $Y$ independent of $X$.

Formally, the problem becomes:
\begin{subequations}
\begin{align}
\min_{p_{Y|X}}{I(X;Y)} \\
\text{s.t.} \sum_{y\in\mathit{Y}}{p_{Y|X}(y|x)U(x,y)} \geq U_c, \forall x \in \mathcal{X} 
\label{eq:constraint0} \\
p_{Y|X}(y|x) \geq 0, \forall x \in \mathcal{X}, \forall y \in \mathcal{Y} \label{eq:constraint1} \\
\sum_{y\in\mathcal{Y}}p_{Y|X}(y|x) = 1, \forall x \in \mathcal{X},\label{eq:constraint2}
\end{align}
\end{subequations}
where (\ref{eq:constraint0}) is
a constraint on the composite utility
$U(x,y)$ defined in \Eq(\ref{eq:com_utility}),
and constraints (\ref{eq:constraint1}) and (\ref{eq:constraint2}) ensure that $p_{Y|X}$ is a valid probability distribution.

We approach this constrained minimization problem by rewriting it as a Lagrange function whose optimal point is a global minimum over the domain of the choice variables and a global maximum over the Karush-Kuhn-Tucker (KKT) multipliers \cite{KKT}. %This approach generalizes the method of Lagrange multipliers to allow for inequality constraints.
We analyze the KKT conditions below to derive key observations on the optimal solution:
\begin{subequations}
\begin{align}
p_X(x)log(\frac{p^*_{Y|X}(y|x)}{p_Y(y)}) - \mu_1^*U(x,y) - \mu_2^* + \lambda = 0 \\
\mu_1^*(U_c- \sum_{y\in\mathcal{Y}}{p^*_{Y|X}(y|x)U(x,y))} = 0, \forall x \in X \\
\mu_2^* p^*_{Y|X}(y|x) = 0, \forall x \in \mathcal{X}, \forall y \in \mathcal{Y} \\
\mu_1^*, \mu_2^* >= 0.
\end{align}
\end{subequations}
Solving this for the optimal conditional probability distribution, we see
\begin{equation}\label{eq:mi2}
    p^*_{Y|X}(y|x) = p^*_Y(y)\exp{(\frac{\mu_1^*U(x,y)+\mu_2^*-\lambda^*}{p_X(x)})}.
\end{equation}
We take the sum of both sides,
\begin{equation}\label{eq:mi3}
    \sum_{y\in\mathcal{Y}}{p^*_Y(y)\exp{(\frac{\mu_1^*U(x,y)+\mu_2^*-\lambda^*}{p_X(x)}})} = 1.
\end{equation}
We then manipulate this to get an expression in terms of $\lambda^*$, which we substitute back into Equation (\ref{eq:mi2}) to get the following:
\begin{equation}\label{eq:mi4}
    p_{Y|X}^*(y|x) = \frac{1}{\eta}p^*_Y(y)\exp{(\frac{\mu_1^*U(x,y)}{p_X(x)})},
\end{equation}
where $\eta$ is a normalization term over $y \in \mathcal{Y}$. From this formal treatment, and reminiscent of our previous work \cite{matt, lillymatt}, we derive two important characteristics of the optimal solution:
%\begin{itemize}
    %\item 
    (i) $p_{Y|X}$ should exponentially increase with utility, and
    %\item  
    (ii) $p_{Y|X}$ should linearly increase with $p_Y$, the probability that $y$ is reported for any $x$, i.e. we should reuse released datasets to the extent practical.
%\end{itemize}

% \subsubsection{Heuristic Approach}
Given the above qualities of an optimal solution, we design the following heuristic approach. We use the pilot dataset to empirically determine the distribution $p_X$ using multi-variate Gaussian kernel density estimation. We then sample from this distribution $N_s$ times to create a ``codebook" which approximates the sample space $\mathcal{Y}$. Limiting $N_s$ allows us to reuse released datasets, as mentioned above.

The weight of each ``code" or possible $y$ value is given by 
\begin{equation}\label{eq:w}
    w(y) = \exp{(\mu_1^*U(x,y))},
\end{equation}
where $\mu_1^*$ is our KKT multiplier. Given an input data $x$, our information theoretic privatizer selects a $y$ from the codebook with probability $w(y)/\sum_{y \in codebook}{w(y)}$. This ensures the likelihood of reporting a $y$ increases exponentially with utility.
As $\mu_1^*$ increases, the IT privatizer offers higher utility but lower privacy. By contrast, as $\mu_1^*$ approaches zero, the IT privatizer achieves lower utility while higher privacy.

In implementation, we use a codebook with size of 51. This codebook size was empirically determined to be large enough that one or more codes would provide good utility, yet small enough that codes are reused to the extent practical. Note that we bias the codebook by including a copy of the unobfuscated data (i.e. 50 obfuscated codes + 1 unobfuscated code). This ensures at least one $y$ has very high utility even for relatively small codebooks. Also, to reduce computational complexity, we split the $n$ measurements into batches and for each batch $x$ we select a batch $y$ from the codebook.

\section{Performance evaluation}
\label{sec:result}

In this section we compare the performance of the privatizers against different adversaries
when users upload a single or a batch of measurements, and
evaluate where they sit in the privacy-utility design trade space.
All performance comparisons in this section are under the typical threat model (bounded adversary).
We use the three real-world traces introduced in Section \ref{subsec:userdata} in our evaluation. Unless otherwise stated, the default trace is the Chania dataset.

%*******************
\subsection{Comparison of Privatizers}
\label{subsec:com_privatizer}

\begin{figure}[t]
\centering
\begin{subfigure}{.23\textwidth}
    \includegraphics[width=.99\linewidth]{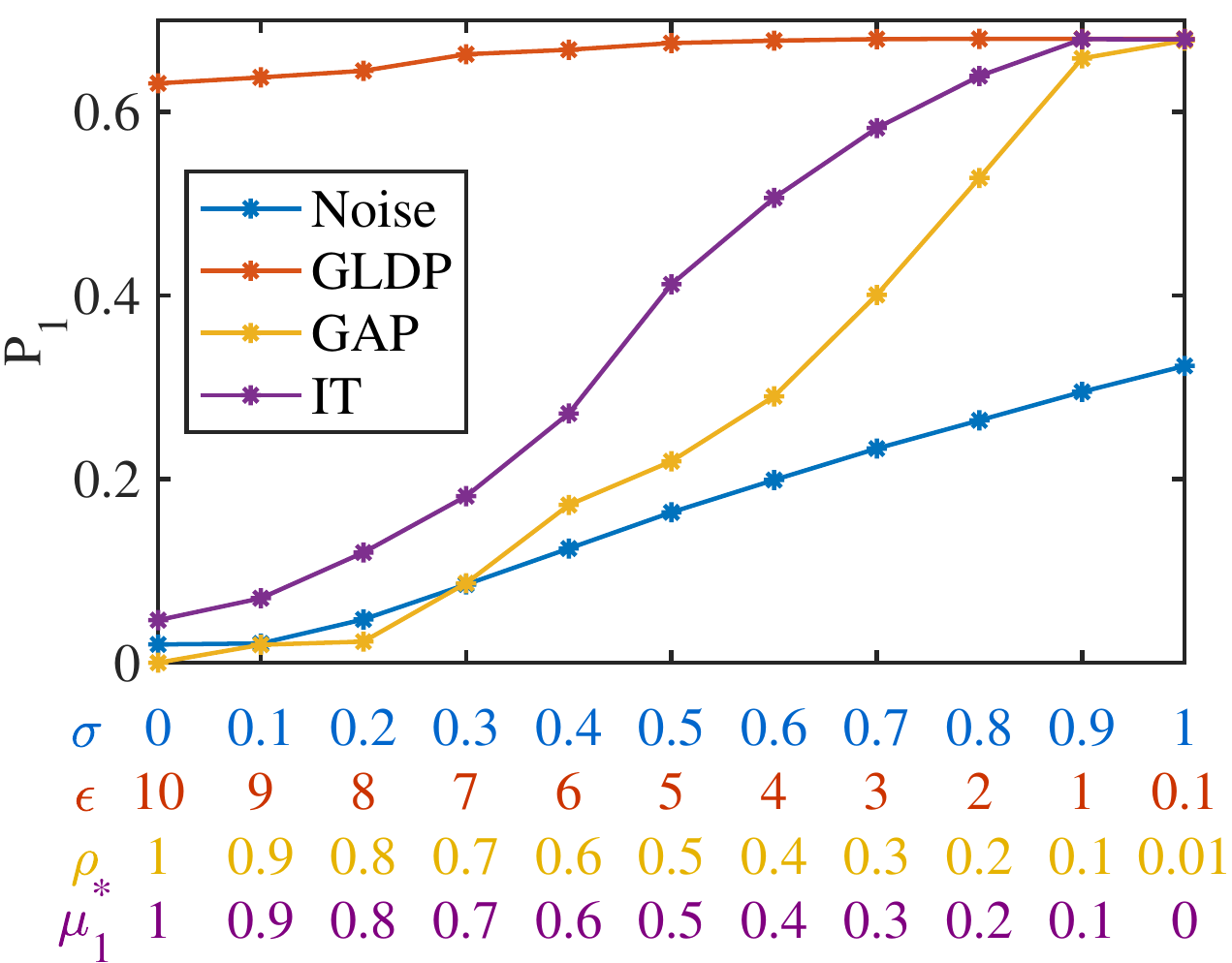}
    \caption{$P_1$: user ID estimate error rate.}
    \label{fig:priv1}
\end{subfigure}
\begin{subfigure}{.23\textwidth}
    \includegraphics[width=.99\linewidth]{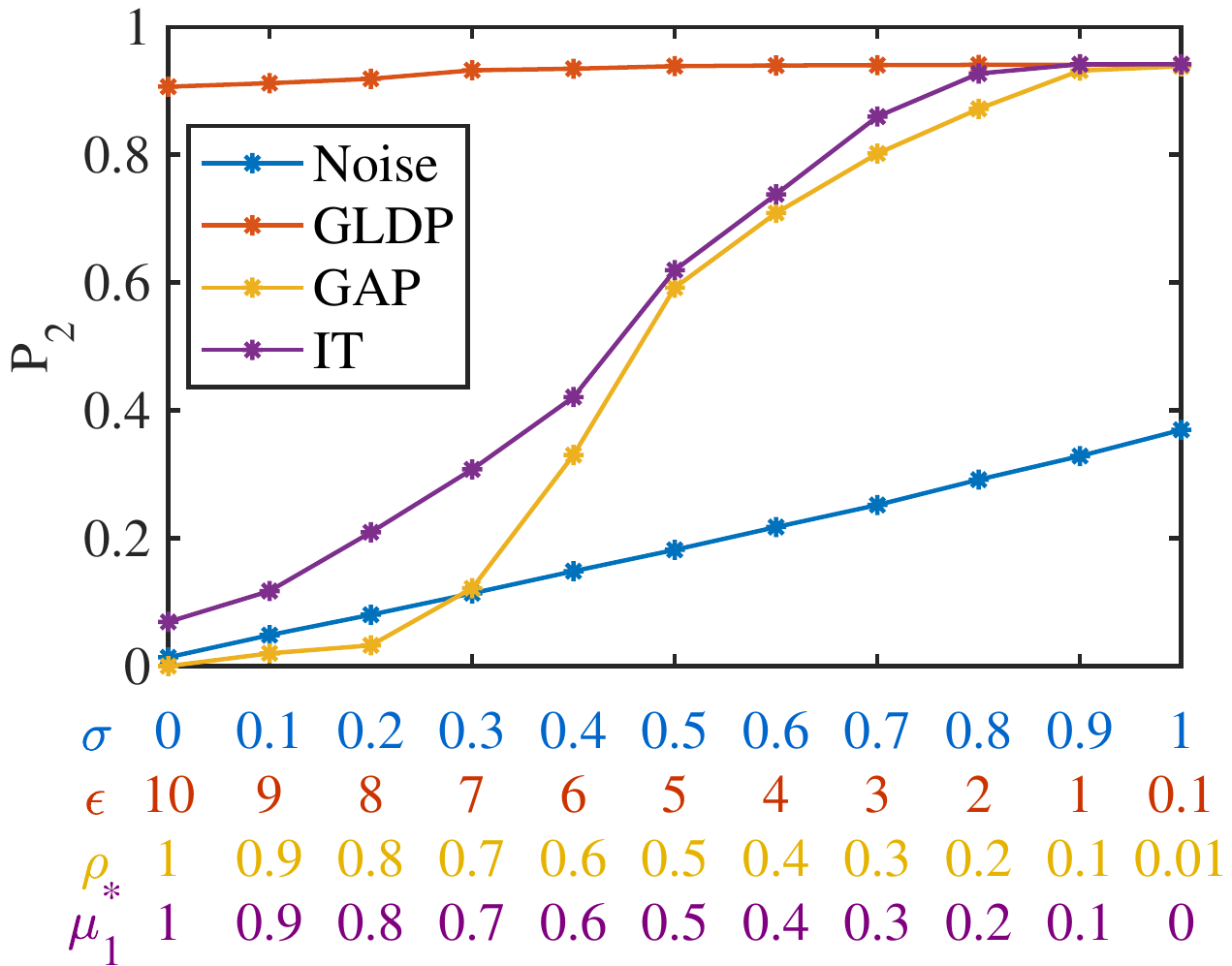}
    \caption{$P_2$: user location estimate error.}
    \label{fig:priv2}
\end{subfigure}
\begin{subfigure}{.23\textwidth}
    \includegraphics[width=.99\linewidth]{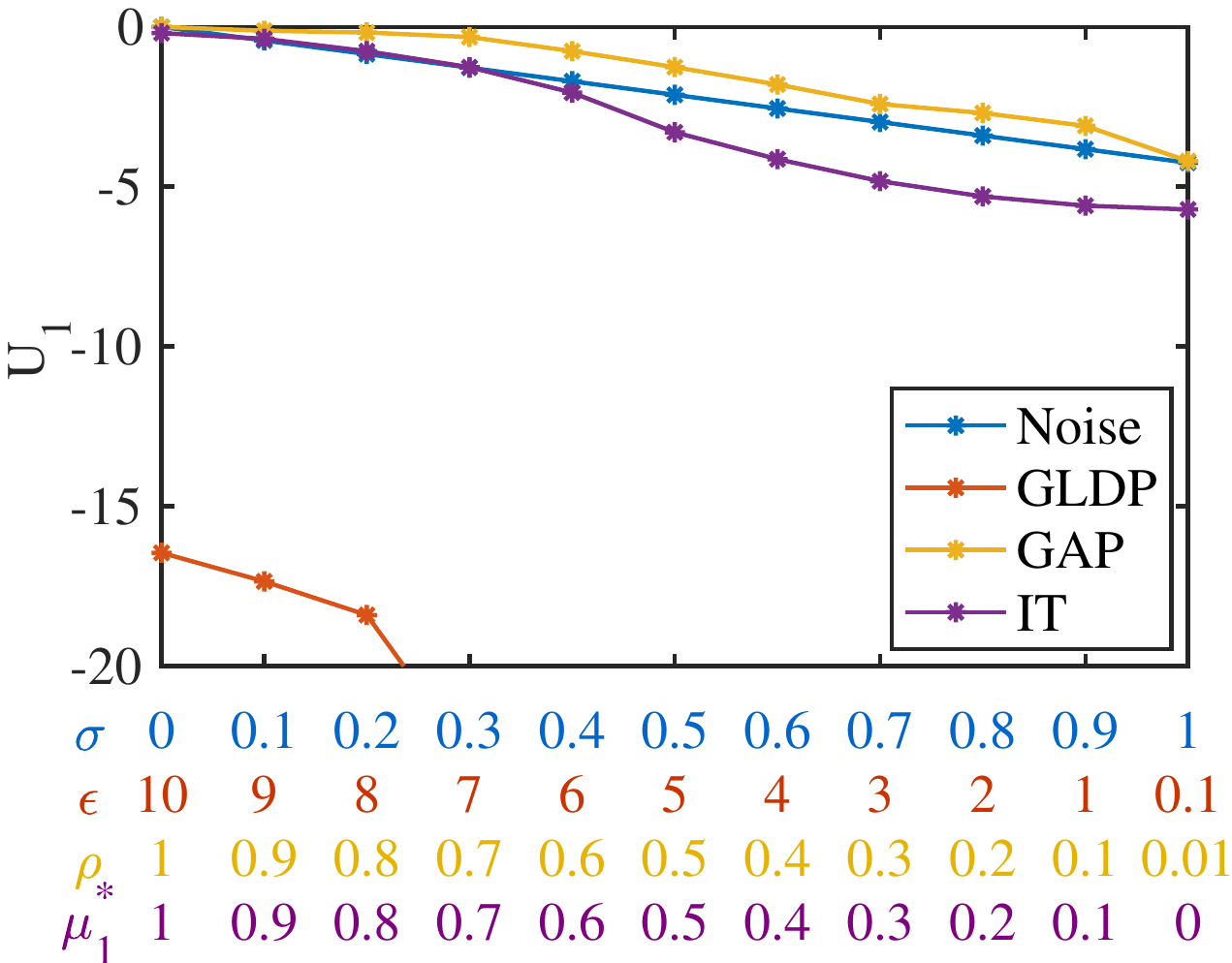}
    \caption{$U_1$: $-$distortion.}
    \label{fig:util1}
\end{subfigure}
\begin{subfigure}{.23\textwidth}
    \includegraphics[width=.99\linewidth]{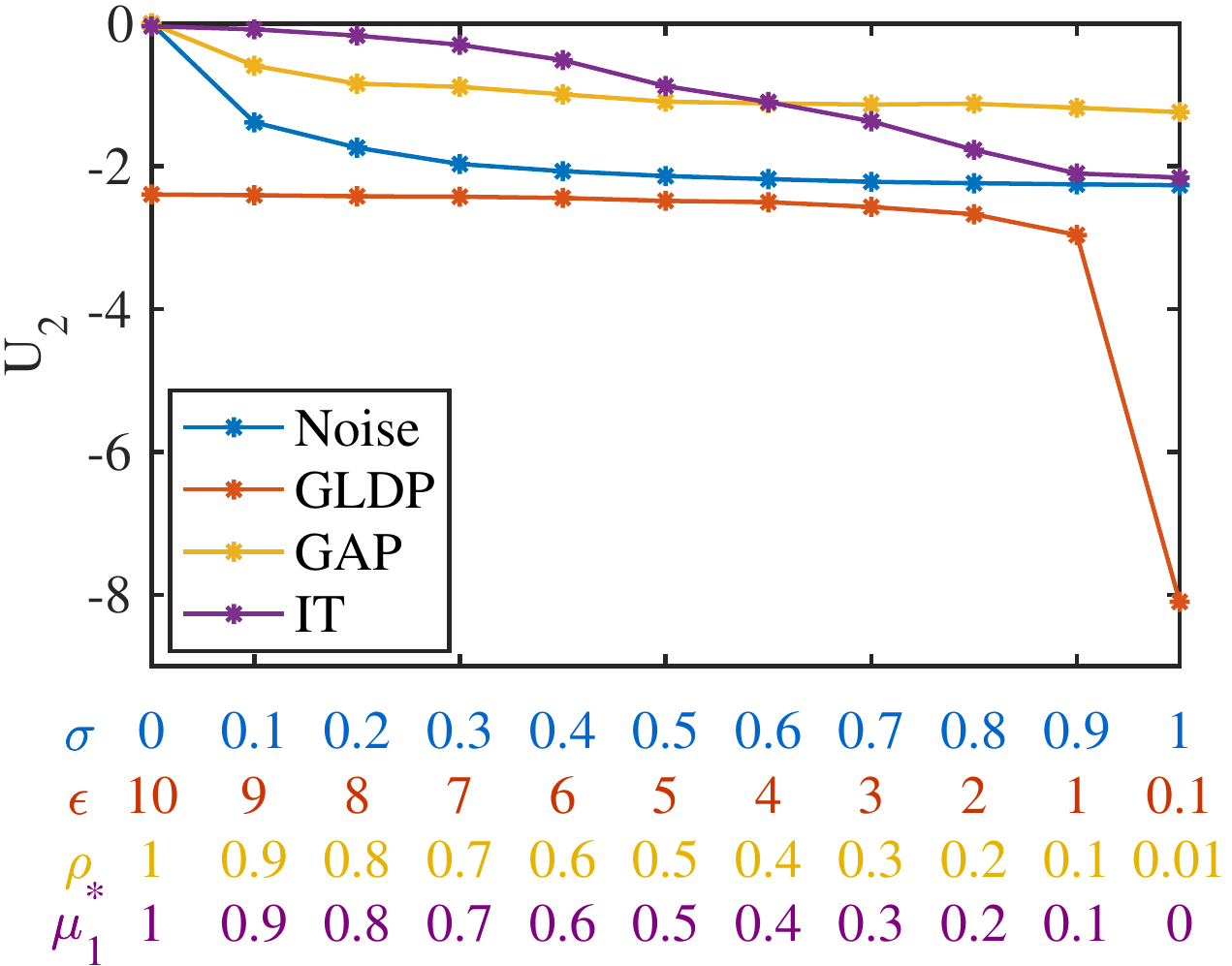}
    \caption{$U_2$: $-$generated map error.}
    \label{fig:util2}
\end{subfigure}
\caption{Privacy and utility of different privatizers. Note that Noise, GLDP, GAP, and IT refer to the Gaussian noise-adding, local Gaussian Mechanism DP, GAP, and the information-theoretic privatizers, respectively.}
\label{fig:util}
\end{figure}

Consider the scenario where users upload a single measurement at a time.
\F\ref{fig:priv1}/\F\ref{fig:priv2} show the adversary estimate user ID error/adversary location error respectively against each privatizer (its privacy), and \F\ref{fig:util1}/\F\ref{fig:util2} show the distortion/generated map error of each privatizer(its utility). The x-axis in these and the following plots represents the parameterization of each privatizer, i.e. $\sigma$, $\epsilon$, $\rho$, and $\mu_1^*$.

As expected, for the noise privatizer, as $\sigma$ increases from 0 to 1 the adversary's user ID and location estimate errors increase, demonstrating higher privacy (larger $P_1$ and $P_2$). At the same time, both the distortion and generated map errors increase, demonstrating lower utility (smaller $U_1$ and $U_2$). For the GLDP/GAP/IT privatizers, decreasing $\epsilon$/$\rho$/$\mu_1^*$ leads to higher privacy (i.e. an increase in the adversary's user ID and location estimate error rate) and lower utility (i.e. an increase in distortion and generated map error).

Among these privatizers, 
the GLDP privatizer consistently achieves high privacy for typical values of $\epsilon$. Specifically, against the GLDP privatizer with $1 \leq \epsilon \leq 10$, the adversary's user ID estimation error is around 70\% and the adversary's location estimate error is close to 1. These numbers can be explained as follows. In the absence of any intelligible patterns due to obfuscation, the adversary learns to assume all measurements came from the geographic center of the dataset, thus its error is on the same order as the spread of input data, i.e. roughly 1.
Both the IT and GAP privatizers can approach this privacy performance as $\mu_1^*$ and $\rho$ get close to 0.1 or smaller.
As for the user ID estimation error, the user with the most measurements contributes roughly 30\% of them, thus a simple adversary assigning this user's ID to all measurements would have 70\% user ID error, hence this can be considered as the upper bound of $P_1$.

With respect to the utility, the GLDP privatizer offers the worst performance. The GAP privatizer outperforms the others for $\rho$ in the range [0.0,0.4] (i.e. high privacy region), while the IT privatizer achieves the best utility for $\mu_1^*$ in the range [0.4,1.0] (i.e. low privacy region). As it will become clear in the next couple of sections, a major reason why GLDP has the best privacy and worst utility is that for the range of $\epsilon$ values considered, it distorts the data to a larger extent than the rest of the approaches. We discuss in more detail the differences between the 4 privatizers in the Section \ref{subsec:tradeoff}. 

\begin{figure}[t]
\centering
\begin{subfigure}{.25\textwidth}
    \includegraphics[width=.99\linewidth]{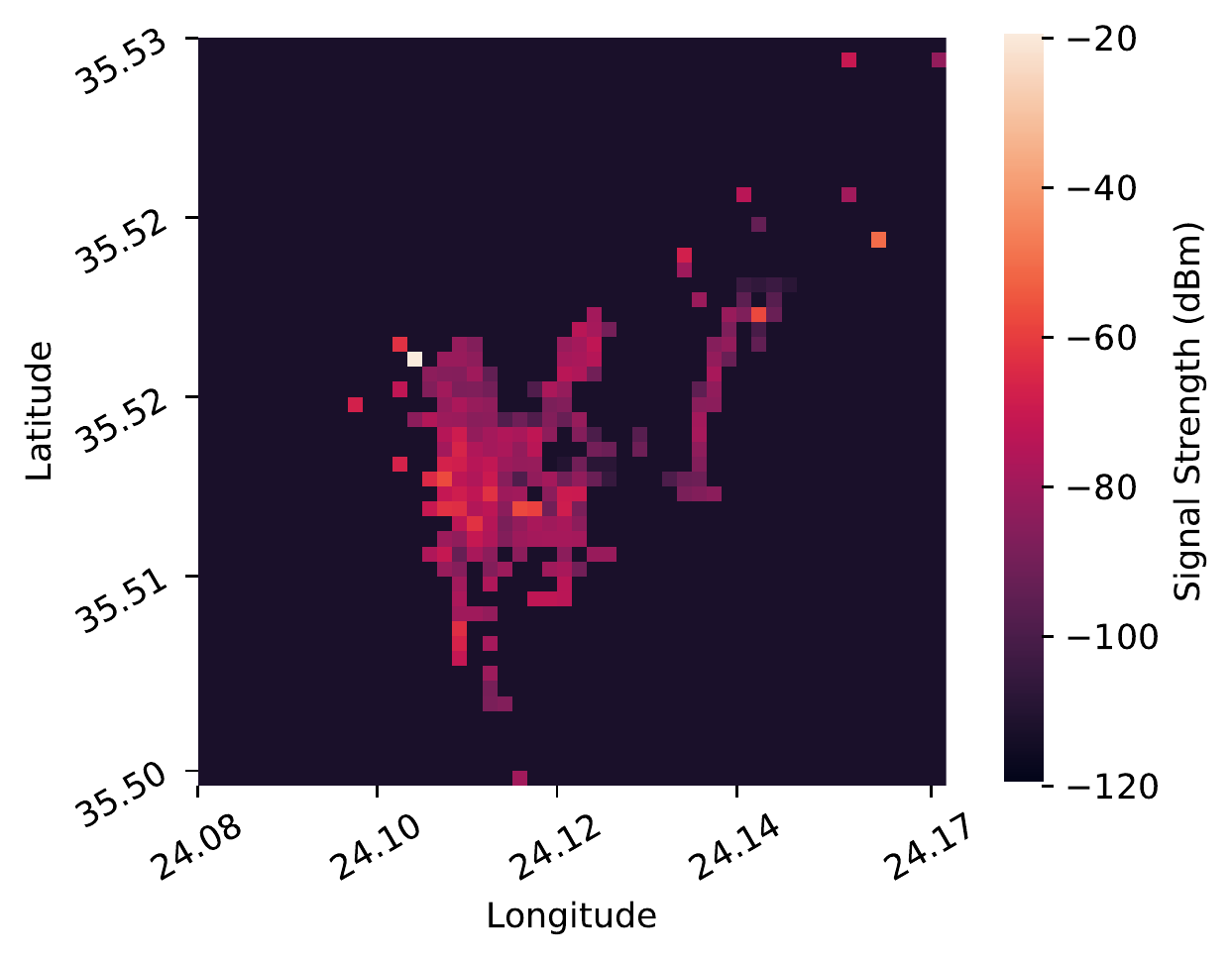}
    \caption{Without obfuscation.}
    \label{fig:map1}
\end{subfigure}
\begin{subfigure}{.21\textwidth}
    \includegraphics[width=.99\linewidth]{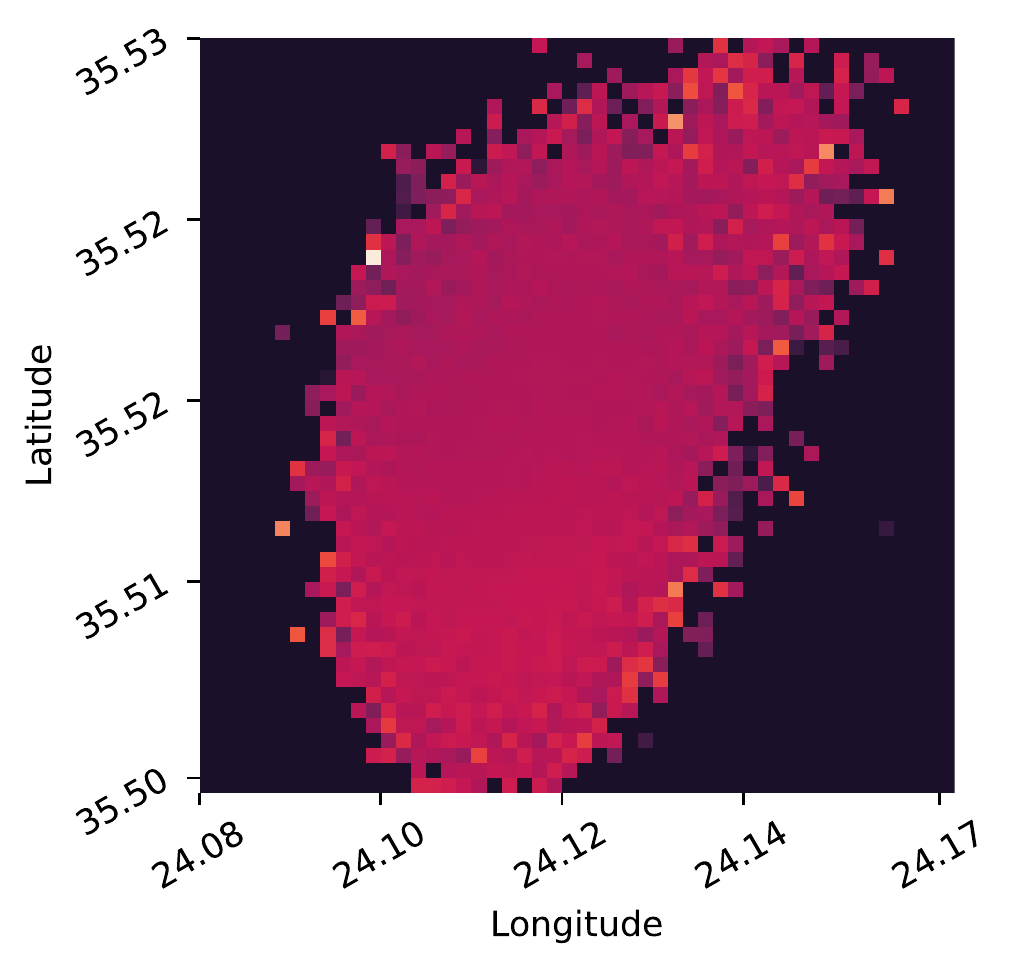}
    \caption{Obfuscated by Noise privatizer.}
    \label{fig:map2}
\end{subfigure}
\begin{subfigure}{.21\textwidth}
    \includegraphics[width=.99\linewidth]{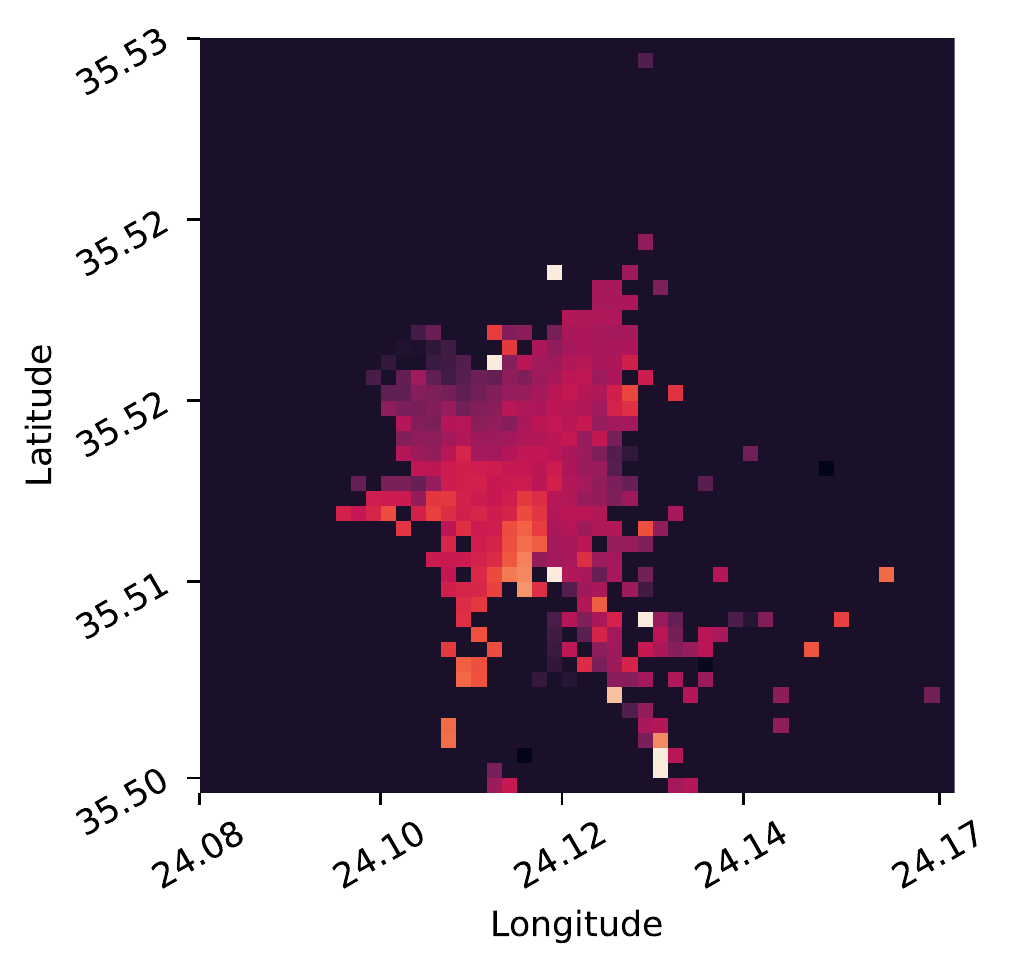}
    \caption{Obfuscated by GAP privatizer.}
    \label{fig:map3}
\end{subfigure}
\begin{subfigure}{.21\textwidth}
    \includegraphics[width=.99\linewidth]{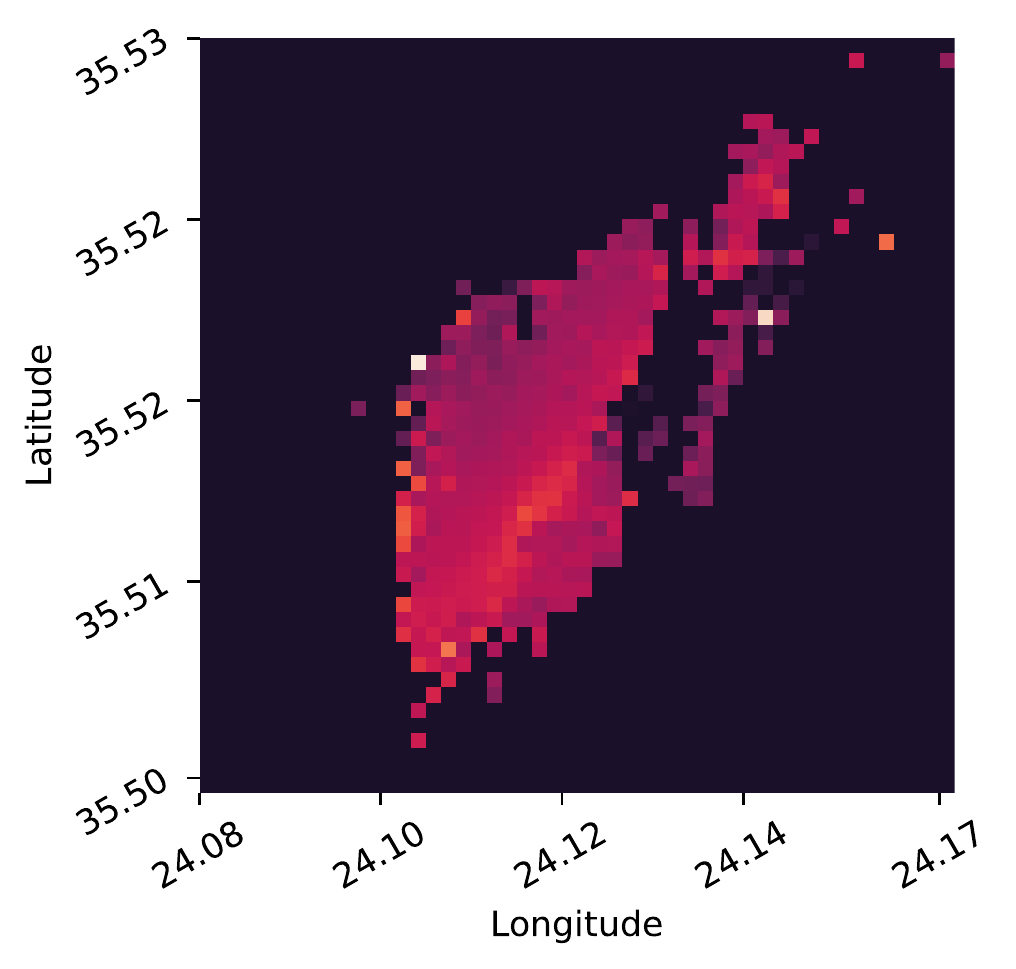}
    \caption{Obfuscated by IT  privatizer.}
    \label{fig:map4}
\end{subfigure}
\caption{Visualization of obfuscated measurements generated by different privatizers when $P=1.0$. Note that each cell in these figures represents a geographical location. The color of each cell represents the average signal strength value in this location. Lighter color represents higher signal strength value. }
\label{fig:map}
\end{figure}

\begin{table}[t]
    \centering
    \footnotesize
    \begin{tabular}{ccccc}
    \hline\hline
         Privatizer & None & Noise & GAP & IT \\\hline
         RMSE (dBm) & 2.46 & 2.90 & 2.54 & 2.48\\
    \hline
    \end{tabular}
    \caption{RMSE (root mean square error) of RSS prediction model trained with obfuscated measurements when $P=1.0$. Note that we use the RSS prediction model trained with non-obfuscated measurements as a baseline (privatizer is none).}
    \label{tab:rmse}
\end{table}

\jiangdraft{Furthermore, we visualize the obfuscated measurements generated by different privatizers as heat maps in \F\ref{fig:map}. Note that the x-axis and y-axis of each heat maps represent longitude and latitude respectively, and the color of each cell in these heat maps represents the signal strength value. Compared with the heat map generated by Noise privatizer, the heat maps generated by GAP and IT privatizers are more similar to the heat map without obfuscation under the same privacy level, indicating that GAP and IT privatizers inject less distortion into the measurement data during obfuscation.}

\jiangdraft{
Lastly, we report the root mean square error (RMSE) of RSS prediction model trained by obfuscated measurements in Table \ref{tab:rmse}. Note that the goal of the service provider is to train a RSS prediction model based on the obfuscated measurements uploaded by users. The more accurate the RSS prediction model is, the higher utility the obfuscated scheme can provide. As illustrated in Table \ref{tab:rmse}, under the same privacy level, the RSS prediction model trained with measurements obfuscated by IT privatizer achieves the lowest RMSE, which is close the RMSE of RSS prediction model trained with non-obfuscated measurements. The RSS prediction model trained with measurements obfuscated by Noise privatizer achieves the highest RMSE, indicating that the Noise privatizer provides the worst utility.
}

% \begin{figure*}[t]
% \begin{subfigure}{.275\textwidth}
%     \includegraphics[width=.99\linewidth]{images/heatmap_noise_0.0.pdf}
%     \caption{Without obfuscation.}
%     \label{fig:map1}
% \end{subfigure}
% \begin{subfigure}{.225\textwidth}
%     \includegraphics[width=.99\linewidth]{images/heatmap_noise_1.2.pdf}
%     \caption{Obfuscated by Noise privatizer.}
%     \label{fig:map2}
% \end{subfigure}
% \begin{subfigure}{.225\textwidth}
%     \includegraphics[width=.99\linewidth]{images/heatmap_gap_0.5.pdf}
%     \caption{Obfuscated by GAP privatizer.}
%     \label{fig:map3}
% \end{subfigure}
% \begin{subfigure}{.225\textwidth}
%     \includegraphics[width=.99\linewidth]{images/heatmap_mi_0.55.pdf}
%     \caption{Obfuscated by IT  privatizer.}
%     \label{fig:map4}
% \end{subfigure}
% \caption{Training data obfuscated by different privatizers when $P=1.0$.}
% \label{fig:map}
% \end{figure*}

\begin{figure*}[t]
\begin{subfigure}{.33\textwidth}
    \includegraphics[width=.99\linewidth]{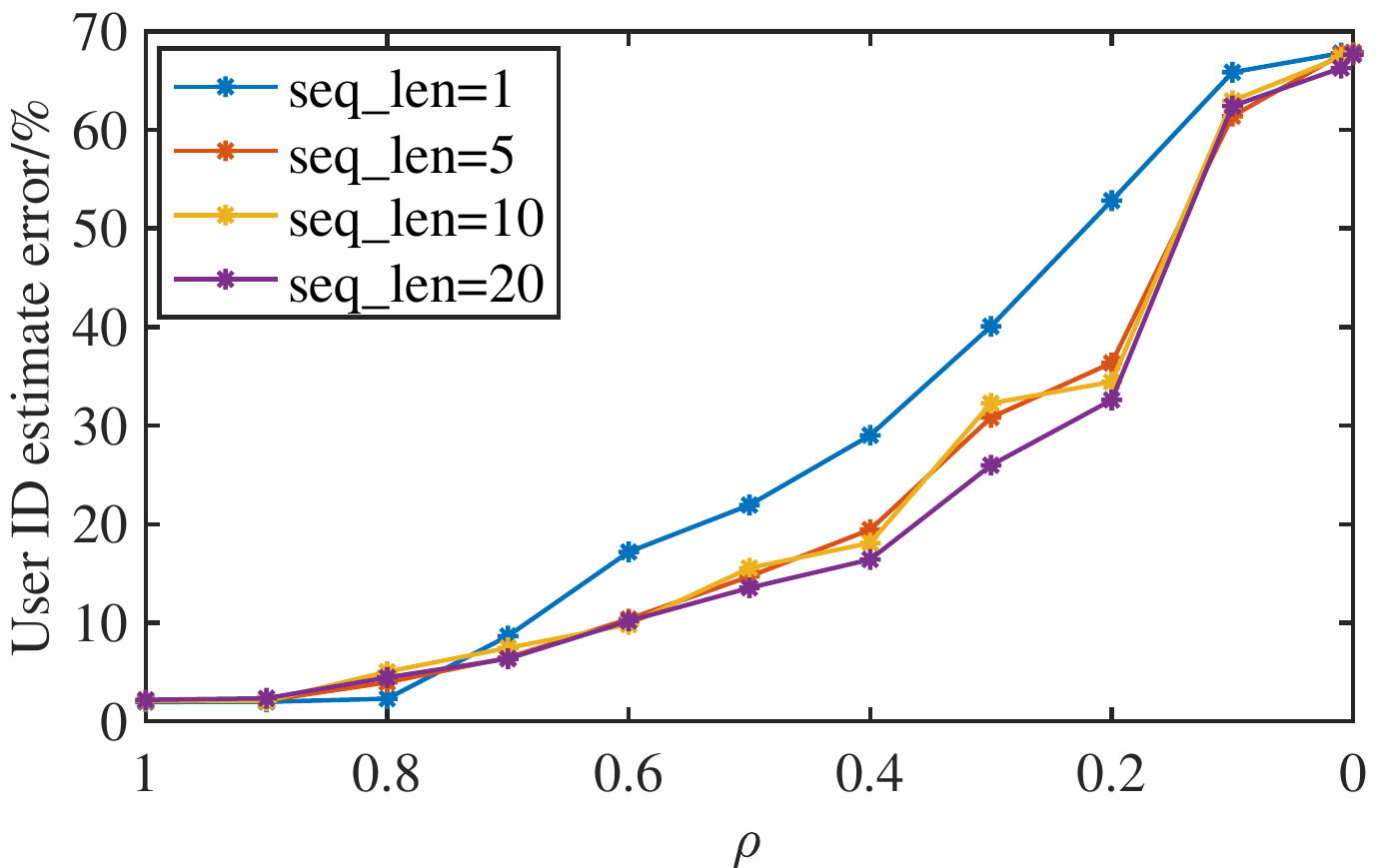}
    \caption{Only adversary leverages sequences.}
    \label{fig:adv_seq}
\end{subfigure}
\begin{subfigure}{.33\textwidth}
    \includegraphics[width=.99\linewidth]{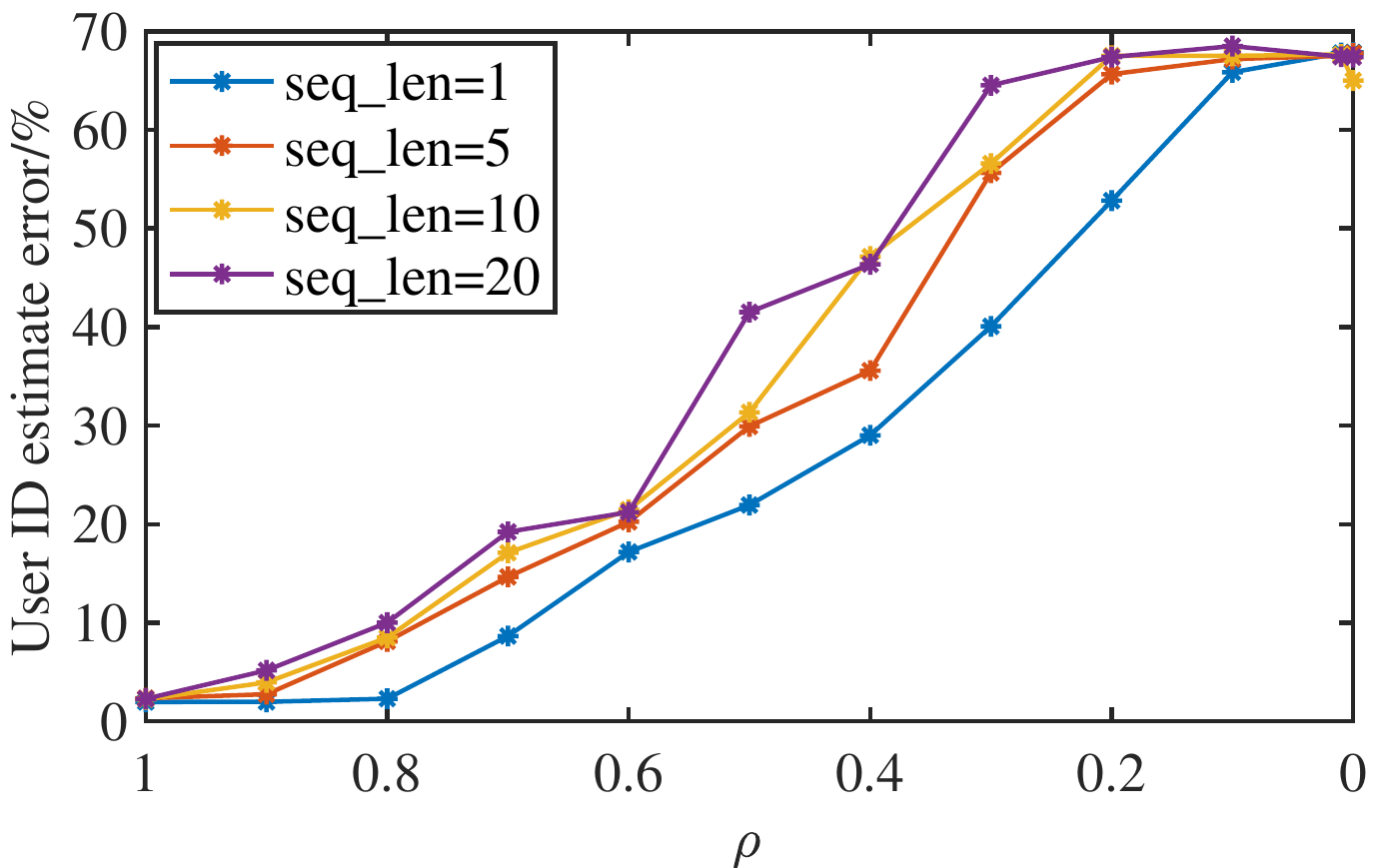}
    \caption{Only privatizer leverages sequences.}
    \label{fig:priv_seq}
\end{subfigure}
\begin{subfigure}{.33\textwidth}
    \includegraphics[width=.99\linewidth]{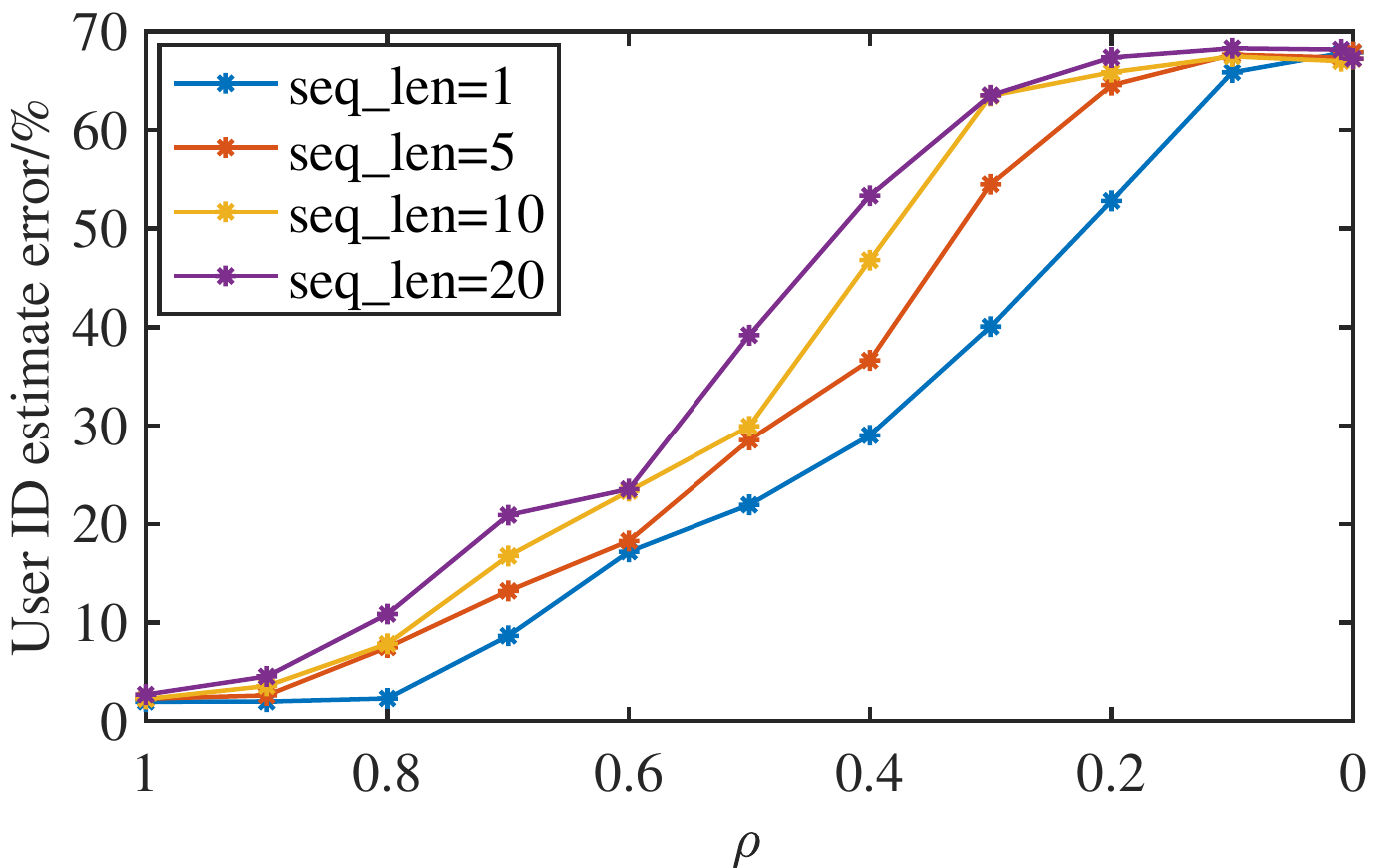}
    \caption{Both leverage sequences.}
    \label{fig:both_seq2}
\end{subfigure}
\caption{Effect of leveraging measurement sequences on adversary user ID estimate accuracy.}
\label{fig:levereging_sequences}
\end{figure*}

\subsection{Leveraging Measurement Sequences}\label{subsec:sequence}

%The adversary modeled in this work thus far does not leverage the fact that mobile data may have predictable patterns when considered as a time sequence. 
To directly investigate the effect of correlations and predictable patterns when considering mobile measurements as a time sequence, we consider an adversary which takes measurement sequences as input, i.e. time sequences of lengths 1, 5, 10, and 20 which belong to a single user, and estimates the (common) user ID of all these measurements, taking advantage of correlations across data of the same user. In practice, the adversary can do this when users upload measurements in batches.

The adversary we consider is trained via supervised learning with the final output of the converged GAP privatizer. The GAP privatizer is a good choice to study sequences of data as it can be trained to consider correlations of sequences and privatize batches of data in one shot as well. 
%(In contrast, say the Noise privatizes data by adding noise in an i.i.d. manner across sequences of data.) 

The results shown in \F\ref{fig:levereging_sequences} consider three cases: only the adversary, only the privatizer, and both of them consider sequences of data.
\F \ref{fig:adv_seq} shows results when only the adversary considers measurement sequences. It shows that the longer the sequence the better the adversary performance, as the adversary achieves smaller error for the same data distortion.
\F \ref{fig:priv_seq} shows results when only the privatizer considers measurement sequences. 
It shows that the longer the sequence the better the privatizer performance, as the privatizer forces the adversary to achieve higher error for the same data distortion. Thus, sequences of measurements help both the adversary and the privatizer, which is expected in the presence of inter-measurement correlations.
That said, the tradeoff in both cases above is the additional computational and memory resources required to handle input sequences as opposed to single measurements.
Lastly, \F\ref{fig:both_seq2} shows results when both the adversary and privatizer consider sequences of the same length. We observe that longer sequences result in better privacy, as the adversary's user ID estimate error increases.

%*********************
\subsection{Analysis of Privacy-Utility Trade Space}
\label{subsec:tradeoff}
\begin{figure*}[t]
\centering
\begin{subfigure}{.32\textwidth}
    \includegraphics[width=.99\linewidth]{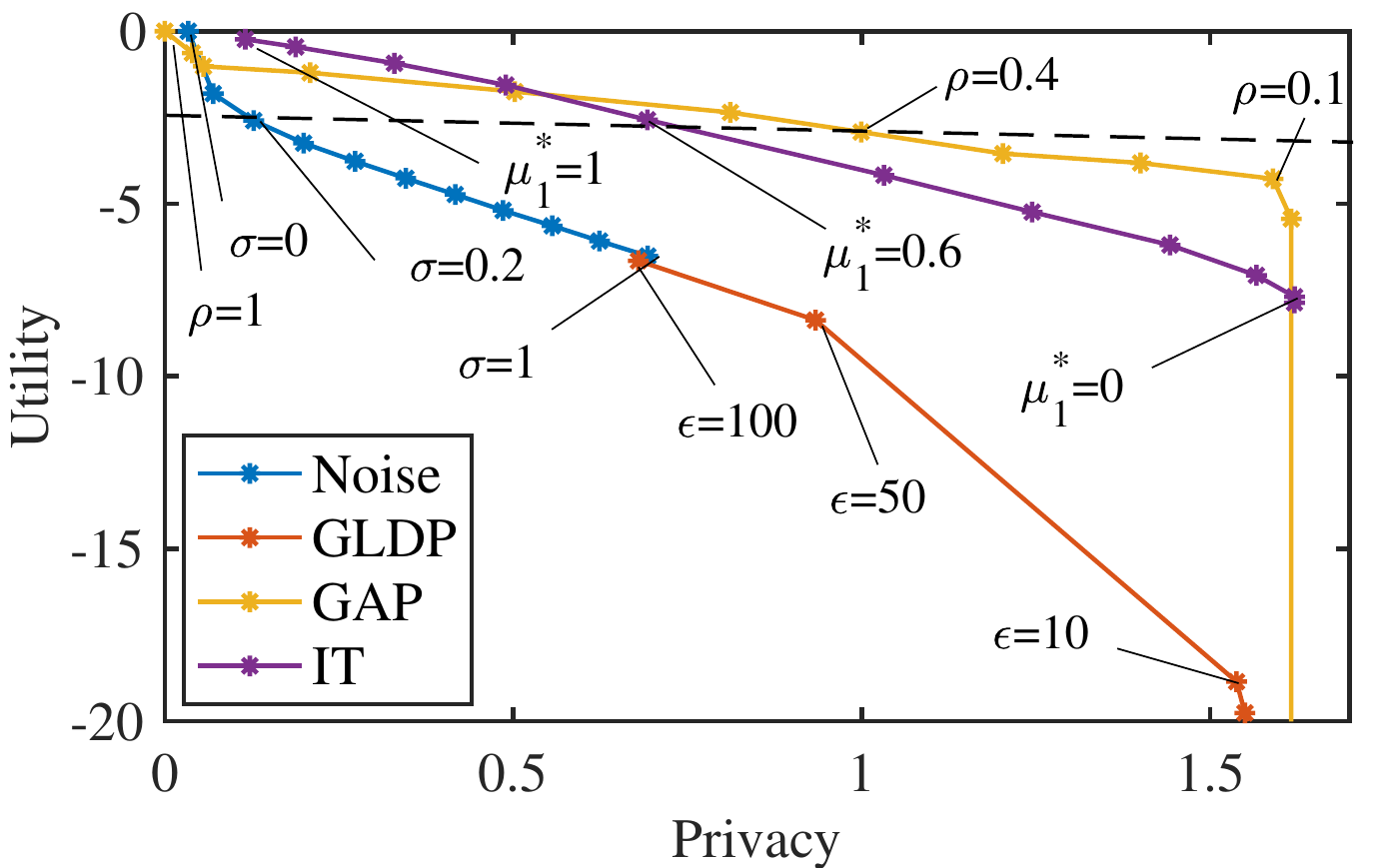}
    \caption{$P$ vs $U$ on Chania dataset.}
    \label{fig:chania_tradeoff}
\end{subfigure}
\begin{subfigure}{.32\textwidth}
    \includegraphics[width=.99\linewidth]{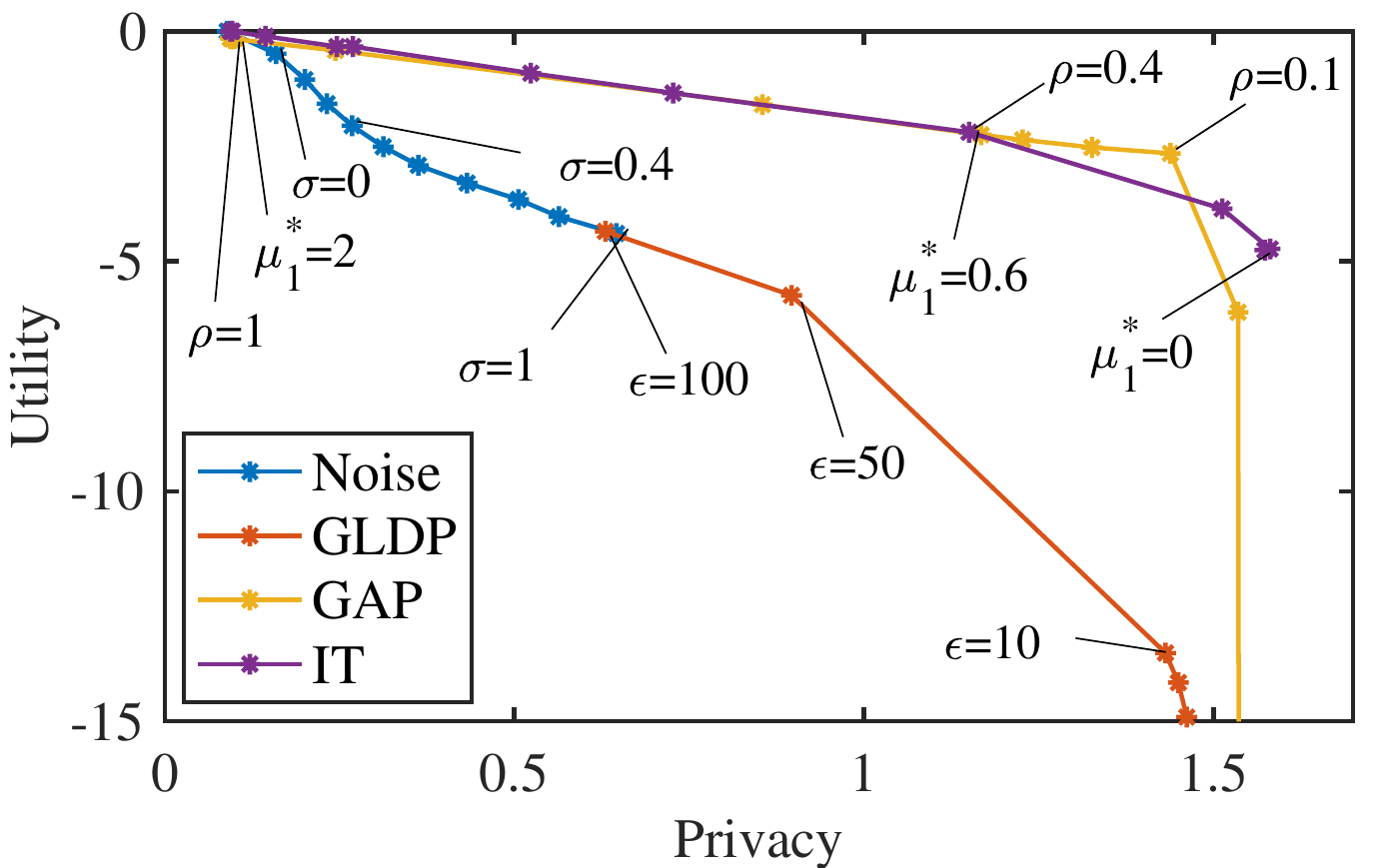}
    \caption{$P$ vs $U$ on UCI dataset.}
    \label{fig:uci_tradeoff}
\end{subfigure}
\begin{subfigure}{.32\textwidth}
    \includegraphics[width=.99\linewidth]{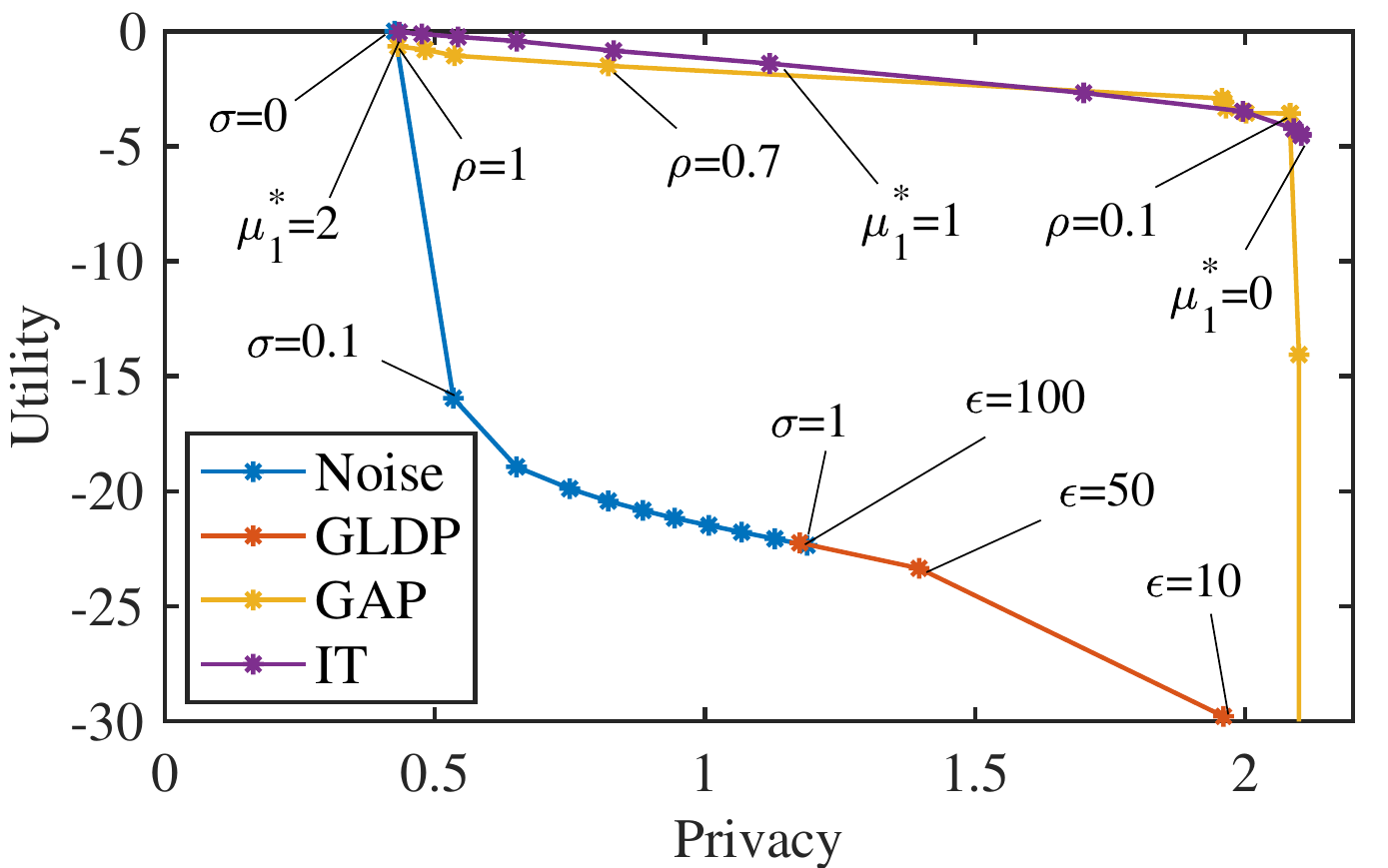}
    \caption{$P$ vs $U$ on Radiocell dataset.}
    \label{fig:radiocell_tradeoff}
\end{subfigure}
\caption{Privacy-utility tradeoff of different privatizers under the Chania, UCI, and Radiocell datasets with composite metrics. Note that Noise, GLDP, GAP, and IT refer to the Gaussian noise-adding, local Gaussian Mechanism DP, GAP, and the information-theoretic privatizers, respectively.}
\label{fig:priv_util_curves}
\end{figure*}

\begin{figure}[t]
\centering
\begin{subfigure}{.23\textwidth}
    \includegraphics[width=.99\linewidth]{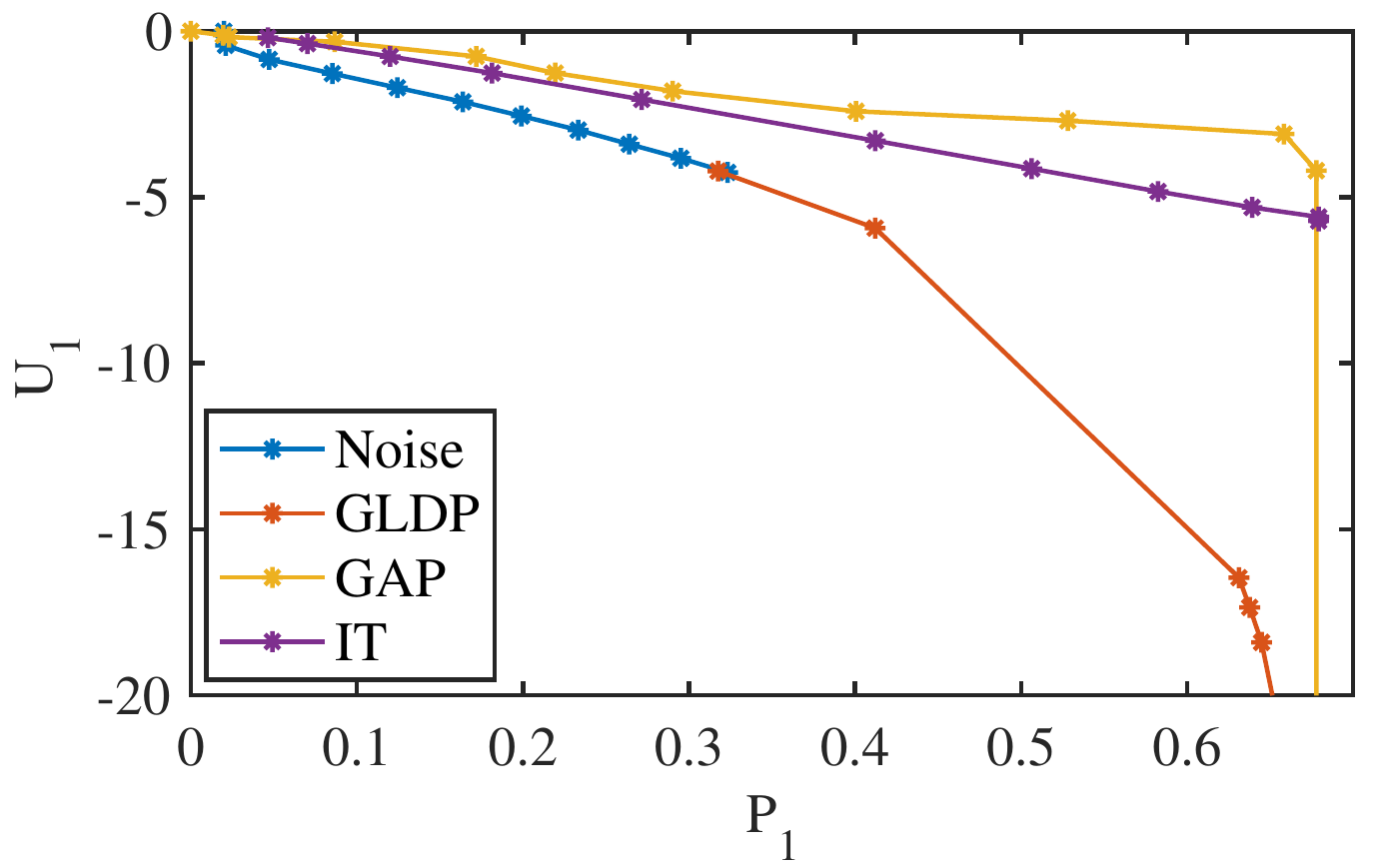}
    \caption{$P_1$ and $U_1$ tradeoff.}
    \label{fig:p1u1}
\end{subfigure}
\begin{subfigure}{.23\textwidth}
    \includegraphics[width=.99\linewidth]{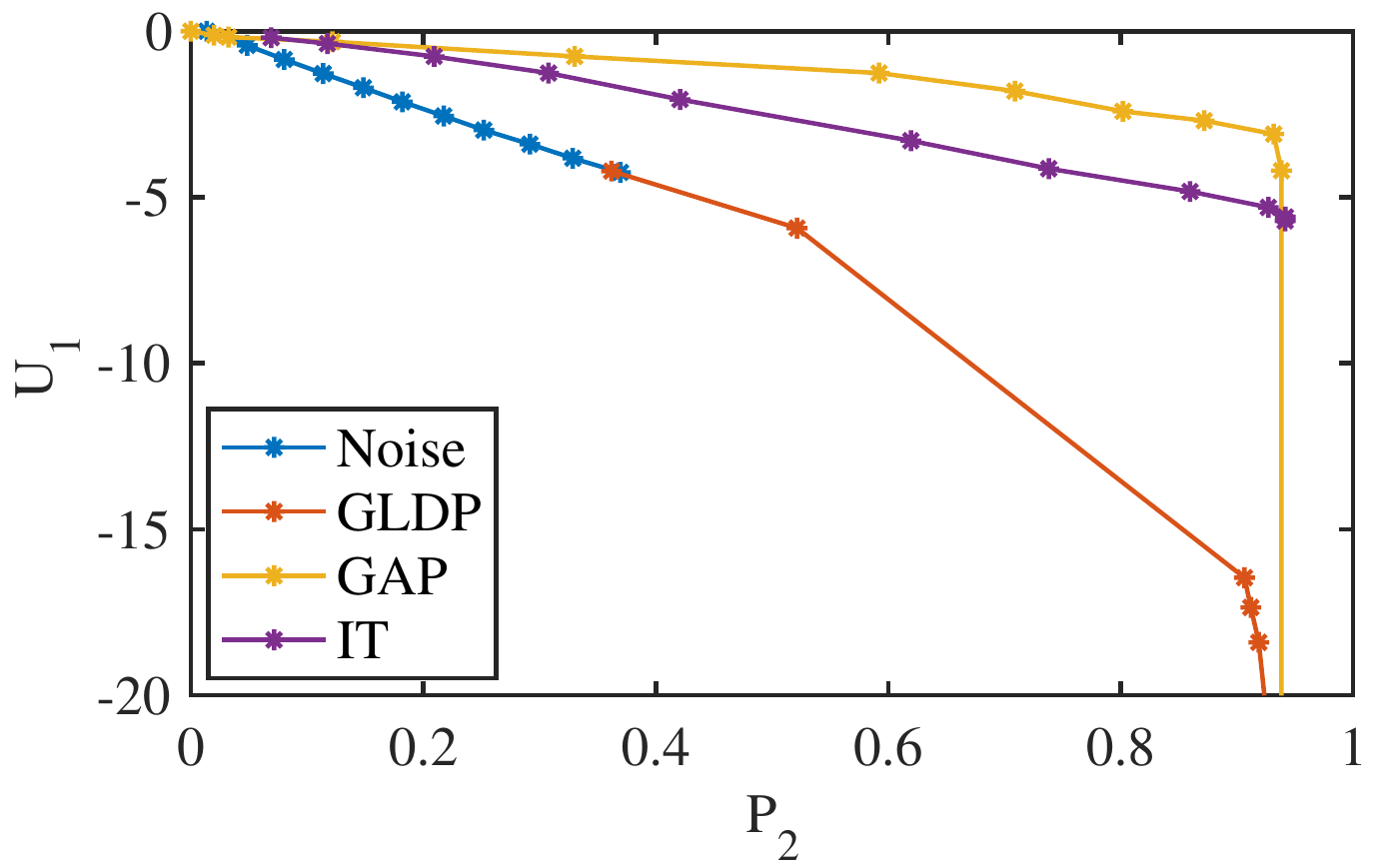}
    \caption{$P_2$ and $U_1$ tradeoff.}
    \label{fig:p2u1}
\end{subfigure}
\begin{subfigure}{.23\textwidth}
    \includegraphics[width=.99\linewidth]{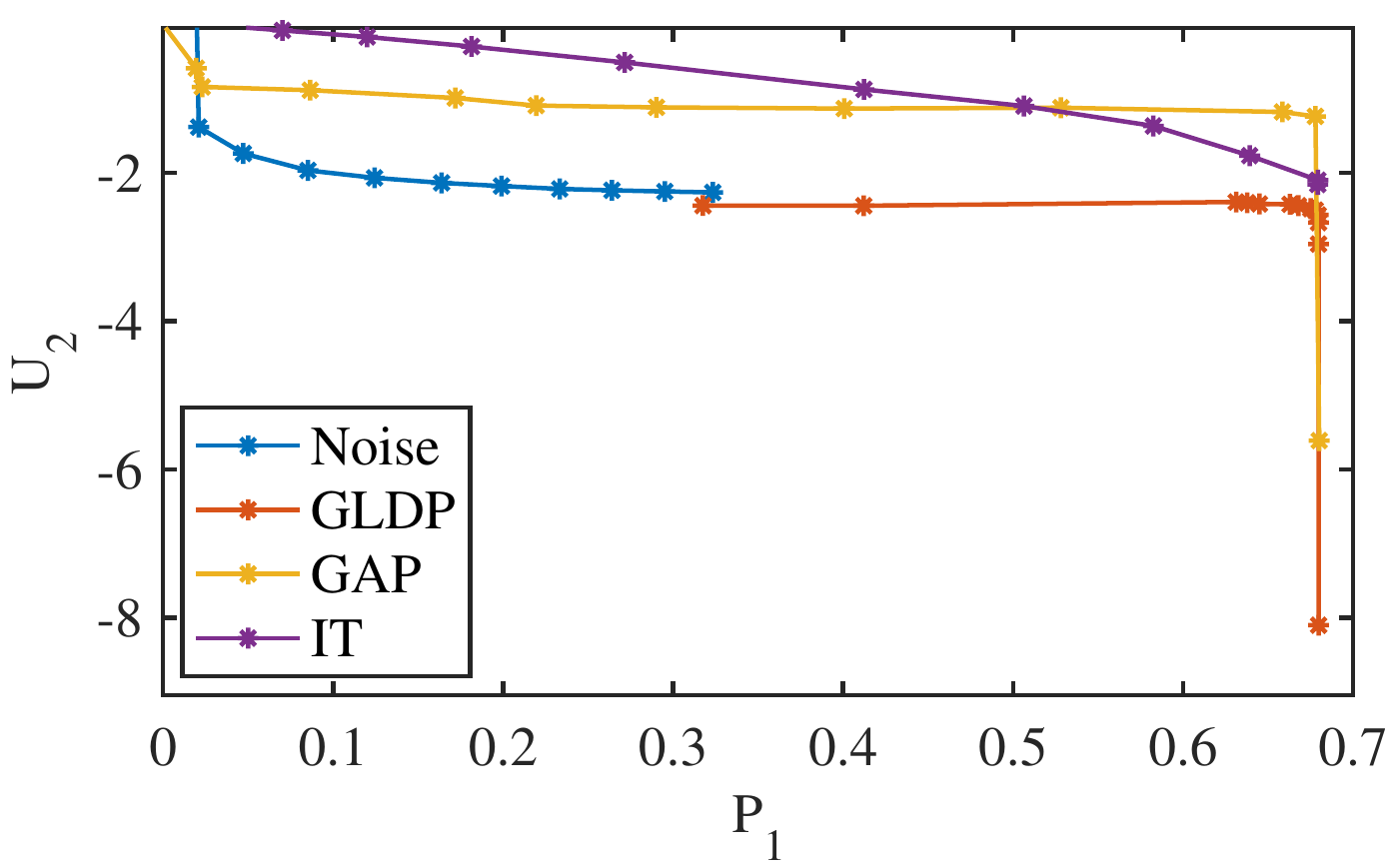}
    \caption{$P_1$ and $U_2$ tradeoff.}
    \label{fig:p1u2}
\end{subfigure}
\begin{subfigure}{.23\textwidth}
    \includegraphics[width=.99\linewidth]{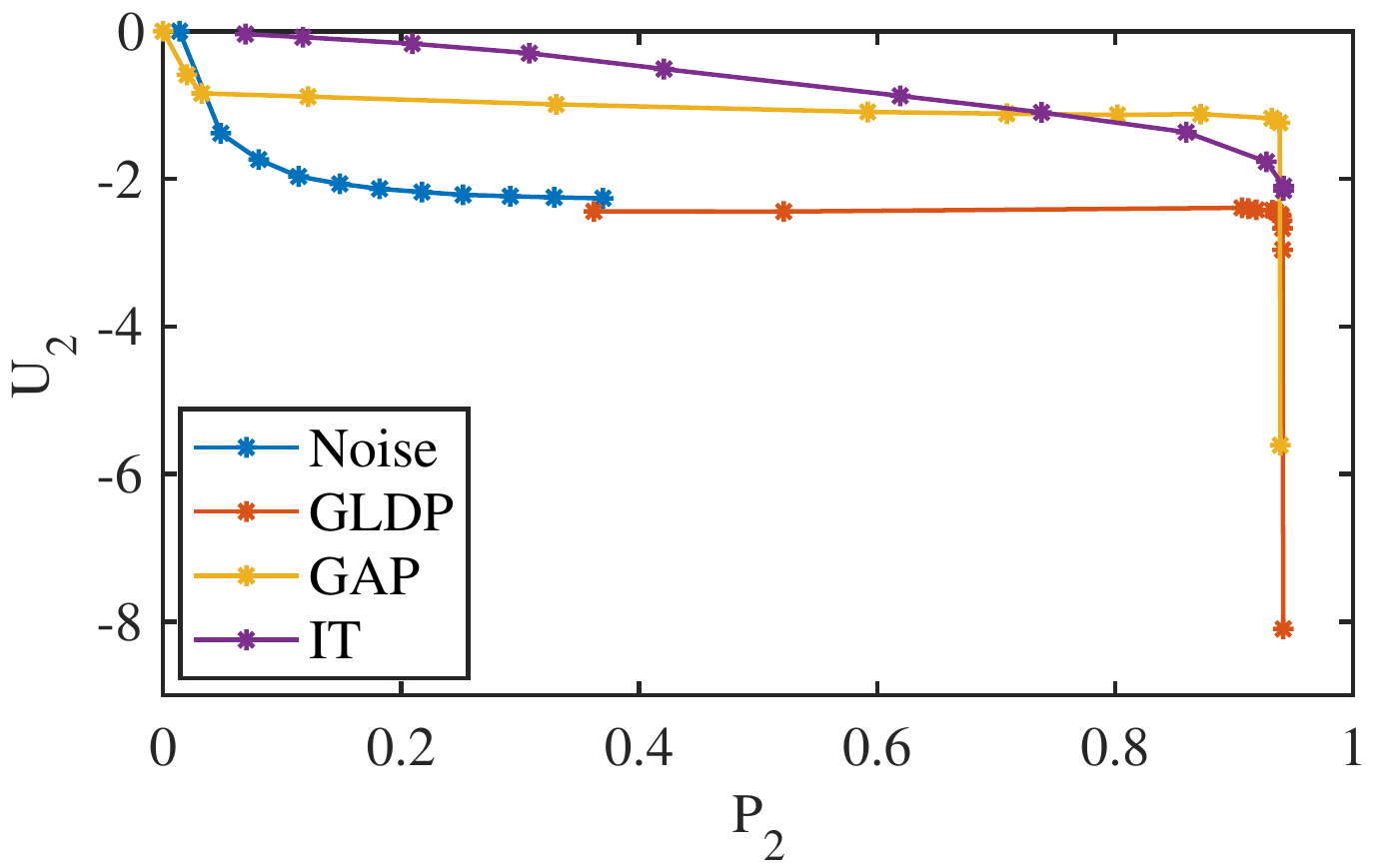}
    \caption{$P_2$ and $U_2$ tradeoff.}
    \label{fig:p2u2}
\end{subfigure}
\caption{Privacy-utility tradeoff of different privatizers under the Chania dataset with non-composite metrics.}
\label{fig:pu_four_tradeoff}
\end{figure}

\F\ref{fig:priv_util_curves} and \F\ref{fig:pu_four_tradeoff}
illustrate where each privatizer sits in the privacy-utility tradeoff space under real-world datasets with different metrics. Specifically, in \F\ref{fig:priv_util_curves}, we consider the composite privacy and utility metrics on Chania, UCI, and Radiocell datasets, where the x axis shows the composite privacy $P$ defined in \Eq(\ref{eq:com_privacy}) with weights $v_1=v_2=1$ and the y axis shows the composite utility $U$ defined in \Eq(\ref{eq:com_utility}) with weights $w_1=w_2=1$. Note that we consider such composite privacy and utility metrics since in practice a service provider may care about both $P_1$ and $P_2$ and about both $U_1$ and $U_2$. 
%we believe the two privacy metrics ($P_1$ and $P_2$) and the two utility metrics ($U_1$ and $U_2$) are equally important for our application.

In \F\ref{fig:pu_four_tradeoff}, we further consider four different combinations of non-composite privacy and utility metrics, and we use the Chania dataset as an example to illustrate how the privacy-utility tradeoff curves of different privatizers change with different non-composite metrics.

In all these plots, the ideal privatizer should sit in the top right corner implying high privacy and high utility.
While the three traces are collected in different countries, areas, and years, the results are qualitatively the same.
From both plots we conclude that GAP and the IT privatizer outperform the Noise and GLDP privatizers. 
It is important to remind the reader that the above comparison is under the typical threat model where the adversary is bounded, whereas GLDP privatizer is the only privatizer that provides privacy guarantees under the worst-case threat model. As discussed in detail in Section \ref{subsec:threatmodel}, we focus on the typical threat model as it is more relevant to our context/application.
%I THINK THE TEXT BELOW IS A REPETITION 
%Going back to the plots, in the very high privacy regime, GAP privatizer achieves equivalent privacy while only sacrificing a quarter of the utility loss GLDP privatizer sacrifices. In the very high utility regime, the IT privatizer achieves equivalent utility with around three times the privacy compared to the Noise privatizer.

%The choice between GAP and IT privacy will depend on which of the two competing goals, privacy and utility, is more important. 

%For applications which require high privacy \textcolor{red}{($P>0.38$ in Chania dataset and $P>0.75$ in Chania dataset)}, GAP will provide better utility. For applications which require high utility \textcolor{red}{($U>-1.36$ in Chania dataset and $U>-2.24 $ in UCI dataset)}, the information-theoretic privatizer will provide better privacy.
%\textcolor{red}{if you address my prior comment the reader will know why you like to compare GAP-DP and InfoTheory-Gaussian and not other pairs :)}

A major reason why GAP and the IT privatizer perform well is that they rely on the notion of \textit{context}, as we have already discussed in Section \ref{sec:context}. The GAP privatizer gains some insight about the structure of the dataset through data-driven learning. It also tries to minimize the difference between the true and obfuscated data while achieving privacy, as encoded in its loss function. In summary, GAP uses $P$, $U$ and the data.
The IT privatizer gains some insight about the structure of the databset through Gaussian kernel density estimation. 
It does well because it releases obfuscated datasets which inherently mirror the true dataset's structure, thanks to a constraint on utility. In summary, IT uses $U$ and the data distribution.  
In contrast to GAP and IT, GLDP only needs information about the data to perform clipping without hurting utility too much (in our implementation we used the data directly for this purpose, see Eq. (\ref{eq:dp4})), and Noise only needs the variance of the data to normalize the amount of Gaussian noise that it adds.

Comparing GAP with IT, because GAP tries to prevent an adversary from estimating features of $x$ given $y$, this strategy can be thought of as a data-driven approach to what IT does, i.e. minimizing mutual information. Yet while the IT strategy adds privacy by choosing $y$ randomly (with appropriate weights), the GAP privatizer maintains a model of a rational adversary which it intentionally tries to deceive. Training against an adversary with the same loss function as the adversary used to test the performance of the privatizers, might be perceived as unfair. To address this, in Section \ref{sec:different_adversaries} we test privatizers against adversaries with different loss functions. 

%\textcolor{red}{We also observe that GAP privatizer outperforms information-theoretic privatizer w.r.t $U_1$ (distortion), while triggering higher value $U_2+U_3+U_4$. The main reason for this is that $U_2$, $U_3$, and $U_4$ are application-specific and only related to location coordinates and RSS value while $U_1$ measures the distortion of all features in each data points. Since GAP is designed to protect privacy feature by obfuscating all the features, intuitively, it would learn to add more noise on features more related to privacy features, while not touching features unrelated to privacy features. Therefore, GAP achieves the best performance on $U_1$ metrics.}

The GLDP privacy-utility curve shown in \F\ref{fig:priv_util_curves} shows values of $\epsilon$ up to 100. Note that this is an order of magnitude greater than the values of $\epsilon$ shown in \F\ref{fig:util1}/\F\ref{fig:util2} and such high values yield a very loose bound on \Eq(\ref{eq:dp1}), yet we do so to show that the Noise and GLDP privatizer meet when noise levels are similar. Note that values of $\epsilon \leq 10$ lie along the asymptotic behavior around the $P = 1.6/1.5/2.0$ line for the Chania/UCI/Radiocell dataset, respectively.

Finally, note that when we train the GAP privatizer and compute the codebook of the IT privatizer to generate the results of \F\ref{fig:pu_four_tradeoff}, we use the composite privacy and utility metrics to avoid retraining/recomputing them for each case. Interestingly, this doesn't deteriorate their performance in a visible manner. While real-world engineers could retrain/recompute the GAP/IT privatizers for the specific privacy and utility metrics they care about, in practice this may be cumbersome.

%*****************
\subsection{Constraining Distortion Levels}
\label{sec:distortion}
\begin{figure}
    \centering
    \includegraphics[width=0.85\linewidth]{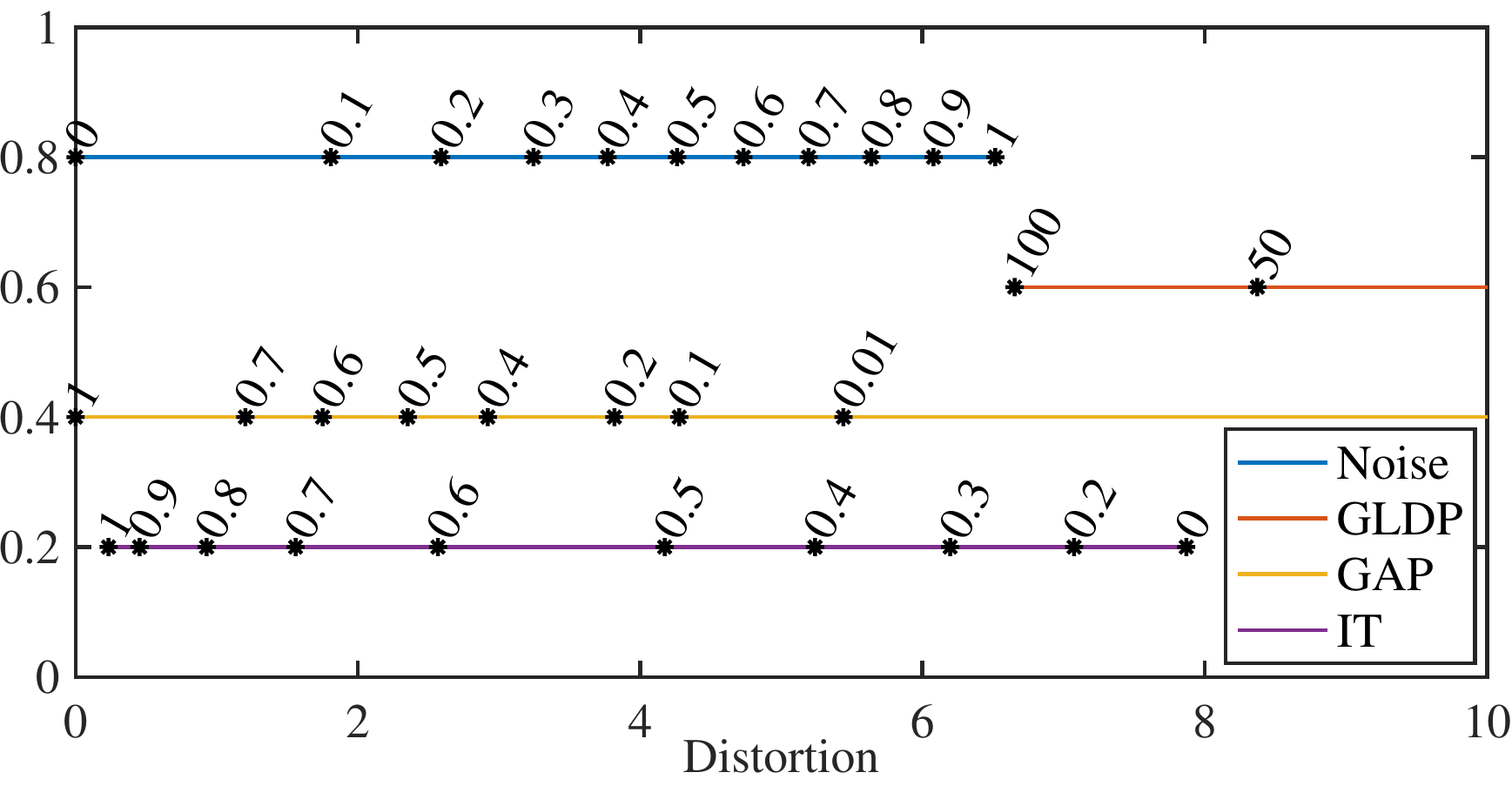}
    \caption{Choosing parameters under a constraint on distortion.}
    \label{fig:approx_eps}
\end{figure}

Previously we have considered privacy and utility as two components of our objective. Suppose instead we wish to maximize privacy subject to a constraint on utility. In \F\ref{fig:approx_eps} we re-frame previous results to demonstrate choosing the appropriate parameters to meet a constraint on distortion ($-U_1$), which can act as an empirical measure of how different the obfuscated data is compared to the original data. \F\ref{fig:approx_eps} presents a plot which can be interpreted as a continuous lookup table. For example, to meet the constraint $-U_1 \leq 3$, we could choose $\sigma=0.2$ or $\mu_1^*=0.6$ or $\rho = 0.4$. 
This plot also offers a sense of which range of distortion each approach may achieve for its selected range of parameter.
%Note that we visualize $\epsilon$ values above $10$ although they are uncommon in the DP literature. 

% The following is an illustrative example of why optimizing privacy/utility subject to a constraint on utility/privacy may be useful. Suppose a network provider would like to purchase data for a competing network provider in order to determine in which areas their competitors have better coverage. The network provider may require that the generated map error be below some value, otherwise they will not pay. For the third party selling the data, the objective may be to maximize privacy (to conform to legislation) subject to this constraint.

%*****************

\begin{table*}[!th]
\centering
\footnotesize
\caption{Evaluation results against common adversaries. Baseline reports the privacy of a privatizer against the adversary trained with its own obfuscated data. Unobfuscated is trained against unobfuscated data.
Aggregate is trained against aggregate obfuscated data. 
Alternative is trained using a different loss function. 
%for all these v1=v2=0.5
}
\renewcommand{\arraystretch}{1.3}
\begin{tabular}{c|c|c|c|c|c|c|c}
\hline\hline
\multirow{3}{*}{Privatizer} & \multirow{3}{*}{Parameter} & \multirow{3}{*}{Utility $U$} & \multicolumn{5}{c}{Privacy $P$ against different adversaries} \\\cline{4-8}
~ & ~ & ~ 
& \multirow{2}{*}{Baseline}
& \multirow{2}{*}{Unobfuscated}
& \multirow{2}{*}{Aggregate}
& \multicolumn{2}{c}{Alternative (different $L_a$)} \\\cline{7-8}
~ & ~ & ~ & ~ & ~ & ~ & $v_1=0.8$, $v_2=0.2$ & $v_1=0.2$, $v_2=0.8$
\\\hline
Noise  &  $\sigma=0.2$  & -2.5
& 0.13   
& 0.30 
& 0.80 
& 0.13  & 0.14\\

GAP  &  $\rho=0.4$ & -2.5    
& 0.95    
& 1.54 
& 1.33 
& 1.05 & 0.88\\

IT  &  $\mu_1^*=0.6$  & -2.5    
& 0.70
& 1.26 
& 1.18 
& 0.68 & 0.70\\ \hline

\end{tabular}
\label{tab:com_adv}
\end{table*}
\renewcommand{\arraystretch}{1}

\subsection{Performance against different adversaries}
\label{sec:different_adversaries}

So far we have tested each privatizer against an adversary trained on the obfuscated data generated by the privatizer. We refer to this as the privatizer's ``own" adversary. What is more, the GAP privatizer is explicitly trained to beat its own adversary and it would be informative to investigate its performance against other adversaries. 

Motivated by the above, we investigate how the four privatizers perform against the following three adversaries: 
(i) ``Unobfuscated" adversary
%adversary \Romannum{1} 
which is trained with the unobfuscated data via supervised learning (rather than the obfuscated data that we have used so far), (ii) ``Aggregate" adversary
%adversary \Romannum{2} 
which has access to all obfuscated data generated by all privatizers, and is trained with the aggregated obfuscated data, and (iii) ``Alternative" adversary trained with a different loss function than the one used so far, which has also been used for training the GAP adversary inside the iterative GAP loop.
Specifically, alternative adversaries use different weights $v_1$ and $v_2$ in the loss function $L_a$.
%Average adversary
%adversary \Romannum{3} 
%which is not an actual adversary but rather corresponds to the following average performance: We position each privatizer against the ``own" adversaries of each of the four privatizer and report the average performance. For example, for the GAP prvatizer we would consider how it does against the Noise, GLDP, GAP and IT adversaries and then report the average. 

Recall that for each privatizer, we have different parameter settings to trade privacy and utility. For a fair comparison, we first set a target utility value and use for each privatizer its parameter value that achieves this utility. 
Table \ref{tab:com_adv} shows the corresponding parameter values for a composite target utility of -2.5. This value is motivated by Table \ref{table:worstcaseutility} and \F \ref{fig:priv_util_curves}, as the former shows the (negative) utility of a random dataset and the latter shows the entire privacy-utility spectrum considered.
(Notice that for GLDP to achieve a -2.5 utility value it would use too large of an $\epsilon$ value ($>$100) thus we omit this line from the table.) We report the privacy achieved by each privatizer against its own adversary (Baseline) and the three adversaries introduced above. 

Interestingly, the GAP privatizer outperforms all the other privatizers not only when privatizers are positioned against their own adversaries (see also Section \ref{subsec:com_privatizer}) but also against the other adversaries, namely Unobfuscated, Aggregate, and Alternative. That said, the performance gap does reduce, which can be explained by the fact that the GAP privatizer is trained against an adversary with a loss function which is now different from that of the adversary used to test the privatizers.

As expected, all privatizers achieve the lowest privacy against their own adversary (baseline), since the latter is trained with the obfuscated data of each privatizer.
Also, all privatizers achieve the highest privacy against the Unobfuscated adversary. This is also expected as the Unobfuscated adversary is trained using unobfuscated data thus it is weaker than the others.
%Another observation is that all privatizers do better against the Aggregate than against the Average adversary.
%\textcolor{red}{This is likely because Aggregate adversary can capture well the complex correlations in the aggregated data consisting of the output of multiple privatizers, making it achieve higher accuracy against different privatizers.}
%GAP and IT  privatizers do better against the Average than against the Aggregate adversary. This is likely because the single ML model of the Aggregate adversary cannot capture that well the complex correlations in the aggregated data consisting of the output of multiple privatizers. 
%Last, as expected, privatizers do better against adversaries trained with obfuscated data of other privatizers than they do against their own adversary. 

\section{Limitations and Future work}
\label{sec:limitation}
%\jiangdraft{In this section, we discuss the limitations in this work, and the future works.}
\noindent\textbf{Points of interest:} The adversary we consider predicts user IDs and all locations from where measurements are collected. However, an adversary may be particularly interested to learn specific users' points of interest (POIs). For instance, the adversary may want to predict the target user's home or work location. We do not consider this in the paper since users can choose to not collect measurements around POIs as a defense mechanism.

\noindent\textbf{Side information:} We assume the adversary has access only to the obfuscated user data shared with the service provider, which does not contain user ID information. A stronger adversary might leverage side information to estimate the user ID of each measurement. For example, the adversary might be able to monitor the network connection between the service provider and mobile users, such that it knows from which device each obfuscated measurement comes from and thus the user ID. This adversary may then build a user whereabouts model. Since it is much harder for an adversary to have access to such information than to merely access database updates, we do not consider
this threat model.

%\jiangdraft{Note that the adversary could also leverage the estimated user ID to build a user whereabouts model. We leave this as future work. We expect that such an adversary would have low accuracy, since a small distortion can already degrade the adversary's user ID estimation significantly in our experiments.}
% Note: Also, given the user ID prediction
% Points of interest. Assume that the adversary has access to the user ID and without shuffling. Hence, it can look at the database, and identify the same points from the same users and get the whole sequence.
% why it is not important in our paper: why these problems are "not" important? If user wants to reduce the prediction accuracy of points of interest, the users can remove them. With respect to non-shuffle user ID, it will not be a problem only if they do it on their own. 
% We leave it as future work. We expect the user ID to be much smaller if there are thousands of users. User ID estimation is not accurate. We expect that in real-world, it will be much less accurate.

% \subsection{Federated learning}
\noindent\textbf{Federated learning:} Mobile crowdsourcing applications lend themselves to a federated learning implementation \cite{kairouz2019advances,li2020federated}, which can provide some privacy for mobile users. Recent works show that federated learning could also leak user privacy \cite{wang2019beyond,nasr2019comprehensive,lyu2020threats,truong2021privacy,FANG2021102199,bakopoulou2021location}. However, it would be a reasonable solution for opt-in mobile users used to collect training data for the GAP privatizer and to estimate data distributions for the IT privatizer.

Another avenue for future work is to investigate how federated learning can be applied, with additional privacy mechanisms, to achieve privacy-preserving training of an RSS predictor. For instance, one may add noise to local model updates or carefully select the measurements used for local model training epochs, to weaken data reconstruction attacks (see, for example, the DLG attack proposed in \cite{zhu2020deep}).

\section{Conclusions}
\label{sec:conclusion}
In this work, we have systematically examined the privacy-utility tradeoff which exists in crowdsourced mobile network data obfuscation.
We have considered four preeminent privatizers employing different obfuscation strategies. 
To compare them, we have identified several privacy and utility metrics as well as a number of adversaries under two different threat models suited to crowdsourced mobile network data, and evaluate the privacy-utility tradeoff performance of different privatizers on three diverse real-world mobile network datasets.
The main takeaway is that under a typical threat model with a bounded adversary, which is of more practical interest in the context of our application, incorporating the structure and intended use of datasets in obfuscation can provide privacy gains without significant utility losses.

%Future work may investigate releasing models learned on data in a private way, federating the adversarial learning to occur on the user's device, or using this framework to test privacy-utility trades in other data obfuscation settings outside of mobile networks.


\section*{Acknowledgment}
We would like to thank Athina Markopoulou and Emmanouil Alimpertis from the University of California Irvine for access to the dataset used in this work \cite{manos}. This material is based upon work supported by the National Science Foundation under grants CNS-1618450 and CNS-1901488.
%% References


\bibliographystyle{elsarticle-num}
\bibliography{./bib/bib}






% \end{thebibliography}

\end{document}